\title{Promises, Impositions, and other Directionals}
\author{Jan A. Bergstra\\Informatics Institute, University of Amsterdam\footnote{%
Science Park 904, 1098 XH Amsterdam, The Netherlands; email \texttt{j.a.bergstra@uva.nl} 
and \texttt{ janaldertb@gmail.com}.}\\~\\
Mark Burgess\\CFEngine\footnote{Email: \texttt{mark.burgess@cfengine.com}.}
}

\documentclass[12pt]{article}

\date{}

\usepackage{alltt}
\usepackage{epsfig}
\usepackage{a4}
\usepackage{times}

\def\beq{\begin{eqnarray}}
\def\eeq{\end{eqnarray}}
\def\2{\frac{1}{2}}

\begin{document}
\newcommand{\promise}[1]{\stackrel{#1}{\longrightarrow}}

\maketitle

\begin{abstract}
Promises, impositions, proposals, predictions, and suggestions are categorized 
as voluntary co-operational methods. 
The class of voluntary co-operational methods is included in the class of so-called directionals.
Directionals are mechanisms supporting the mutual coordination of autonomous agents.

Notations are provided capable of expressing residual fragments of directionals. An extensive example,
involving promises about the suitability of programs for tasks imposed on the promisee is presented. The example
 illustrates the dynamics of promises and more specifically the corresponding mechanism of trust updating and
credibility updating. Trust levels and credibility levels then determine the way 
certain promises and  impositions are handled. 

The ubiquity of promises and impositions is further demonstrated
with two extensive examples involving human behaviour: an artificial example about an agent planning a 
purchase, and a realistic example describing technology mediated interaction concerning the solution of
pay station failure related problems arising for an agent intending to leave a parking area.

\end{abstract}
\newpage
\tableofcontents
\newpage
\section{Introduction}
The objectives of this paper are diverse, including the following:
\begin{enumerate}
\item to discuss in some detail the dynamics of 
promises assuming the statics of
promises to have been assessed to some satisfactory extent in \cite{BergstraB2008}.\footnote{%
The role of autonomy for agents acting upon the reception of decision 
outcomes has not been brought into focus in my 
work \cite{Bergstra2011a} on outcome oriented
decision taking and in these papers the possibility that decision outcomes imply 
obligations for other agents is left open.}

\item To add impositions to the theory of promises thus achieving higher symmetry. 
An imposition is an invitation for voluntary cooperation.

Impositions are most plausible in contexts where the target agent of an imposition has in 
advance promised the corresponding source of the imposition of its willingness to 
receive and to subsequently effectuate a sufficiently large class of impositions.

\item To add proposals, predictions, suggestions, and warnings as methods for inducing voluntary cooperation 
similar to though different from promises and impositions. 

Incorporation of these further methods allows for a more flexible
application of promises and impositions in human management and organization. 
For instance predictions play a role when knowledge about an environment must be shared between 
different agents. Suggestions and proposals are
exchanged during preliminary stages in advance of an exchange of promises and impositions.

\item To collect promises, impositions, proposals, predictions, suggestions, and warnings into a category of so-called
co-op (short for co-operational) methods which share to a large extent options for formal 
description as well as life-cycle models and
method dynamics. Penalties for non-compliance play no role in the setting of 
voluntary co-operational methods. Below voluntary co-op methods will be referred to as ``directionals''. 
Directionals constitute a larger class of methods for achieving coordination between autonomous agents including messages, hints, 
smiles, outcries, alarms.

\item To provide examples of promises and other directionals that demonstrate the interaction 
with trust maintenance which in turn is the key to promise dynamics. And to collect a number of 
additional attributes of promises that are helpful for an understanding of promise dynamics.
\end{enumerate}

We follow the initial development of \cite{Burgess2005,Burgess2007,BurgessF2007} for an approach to 
a theory of promises with a principled emphasis on agent autonomy. 
A simple notation for promises, involving four components:   a promiser, a promisee, 
a promise type, and a promise body, provides remarkable expressive power. A range of examples 
has been put forward that demonstrates the
virtue of the use of promises for system specification and understanding. 

In \cite{BergstraB2008} that work has
been extended in the direction of providing conceptually richer definitions of core concepts. In \cite{BergstraB2008}
no attempt was made to specify dynamic aspects of promises except for the general observation that a promise
will be assessed by agents in its scope. The idea of promise keeping is considered implicit in the outcomes of assessments rather than to be recovered from some objective form of observation or even logic.

Contemplating promise dynamics below will lead to the identification of a number of features that
may be attached to promises on top of the four components mentioned above. This in turn leads us to the
specification of
a variety of promise statement notations extending the expressive power of the base
notation from \cite{Burgess2005,Burgess2007} and \cite{BurgessF2007}.

At a closer inspection promise dynamics, as well as imposition dynamics, and the dynamics of other directionals
is dominated by levels of mutual trust between agents and the design and implementation 
of trust maintenance functionality. This is
illustrated in an extensive example about the usage of computer programs in relation to the trust level of a consultant
advising other agents about the use of that software.

In Appendix~\ref{appendixA} an example is provided which 
indicates that for a simple plan involving a few actions only a
multitude to promises, counter promises, proposals, and suggestions may be issued. 
This example illustrates the importance of promises for plan formation. 
In Appendix~\ref{appendixB} an extensive case study is provided which illustrates that promises may be useful
for the explanation of service architectures and of difficulties arising during service delivery. The use of promises 
seems to be unavoidable in this case, which increases our confidence in the role that promises may play
in the description of systems involving human-machine interaction.

\subsection{Liberating promises from the connotation of obligation}
A design decision that underlies promise theory is to liberate the concept of a promise from the
connotation, or implicit expectation, that a promise correlates one to one with an obligation. 
In \cite{BergstraB2008} several arguments have been put forward why that so-called non-obligationist 
conception of promising may be of practical value, both inside and outside computing.

The main reason for disentangling promises from obligations is that in a world of autonomous 
agents promising is unproblematic, whereas obliging is not. One agent imposing an obligation
on another agent may be understood as an impairment of the second agent's autonomy. 
This objection against promises being strongly coupled with obligations depends on a conception 
of obligations that may be questioned. Although the decoupling of promises and obligations 
has been dealt with extensively in \cite{BergstraB2008} we feel that more ought to be said. Below we will
outline how some promissory obligations may be understood as bundles of promises.

\subsection{Impositions and impository obligations}
Another common source of obligations arises when one agent commands another to perform some action which
the ``must be'' performed. However, just as with promises one may remove the connotation of obligation
from a command. We will speak of an imposition instead.  An imposition is
issued by an impositioner to an impositionee. 

Unfortunately ``imposition'' has a negative connotation because an imposition is generally understood to
be be unwanted by its target agent. Nevertheless there is a striking congruence between impositions and
promises because source, target, scope, type, and body, each make sense in similar ways.

Just as to each promise a promissory obligation can be found which grasps what is obliged to the promiser
(which may be nothing), one may find an impository obligation for an imposition which collects that what becomes
an obligation to the impositionee upon the imposition being issued. In many cases the impository obligation is empty.

\subsubsection{Liberating impositions from the connotation of unfairness}
\label{Imp}
After undoing ``imposition'' from its connotation of unfairness, if only in the context of promise theory,
an imposition becomes a symmetric counterpart to a promise. In order to have a perfect symmetry the
preferable interpretation is thus:
\begin{enumerate}
\item Promise: act of promising, event of promise issuing, resulting in a promise outcome,
\item Promise outcome: essence, or merely description, of what has been promised. The promise outcome 
is specified as a component of a promise statement in the so-called promise body.
A promise outcome equals an imposition enacted by the promiser towards itself.
\item Imposition: the result of an act/event of imposing. An imposition may be either external
or it may be self-imposed (a self-imposition created upon a promise),
\item Imposition event/action: source agent tells target agent what to do, 
what to achieve, what must be
the state of affairs the the target must see to that is reached.
\end{enumerate}


\subsubsection{Impositions as a tool for autonomous agents}
Now the vital step is to mobilize an ``imposition'' for interaction between autonomous agents. This requires 
a number of assumptions:
\begin{itemize}
\item There is no underlying hierarchical structure that explains or governs who may impose on whom.
$A$ may say to $B$, ``please open that door for me'', and that can happen for all agents $A$ and $B$ 
(assuming these agents deal with doors). 

\item An imposition of $A$ on $B$ should not be understood as an attack by $A$ on $B$. 
Rather an imposition constitutes an attempt by $A$ to induce voluntary cooperation for a 
certain objective or course of events from $B$.
\item Suppose that $A$ imposes $p$ to $B$, then $B$ may degrade its respect for $A$, 
respect being an additional status feature besides credibility and trust. If $B$ fails to comply with $p$
then $A$ may degrade either its respect or its trust of $B$. And conversely if $B$ subsequent behavior achieves $p$.
\item $B$ may be happy to comply with an imposition $p$ issued by $A$. For instance if $A$ says
to $B$: 
``our chairperson $C$ is delayed, and you chair the opening 
session of this conference now'', then $B$ might be honored 
and very willing to do so. $B$ might also be embarrassed, there is no way to tell in advance.
\item For an agent $B$ the collection of all open impositions (as were issued by any agents including
$B$ by way of its promises), represents a to do list which $B$ can act upon, depending on its own
preferences, which take into account the impact of $B$ actions 
on other agent's respect and trust for $B$, as eel as $B$'s reputation in general.

\item $A$'s issuing an ``unfair'' imposition on $B$ is reflected by, potentially a decrease of respect
and/or trust in $A$ from $B$, and from other agents in scope of the act of imposition. 
There is no need
to have a definition or description of fairness or unfairness other than what 
may be derived from how different agents update their trust and respect of 
$A$ upon the imposition being issued. Promise and imposition are equally neutral.
\end{itemize}

\subsubsection{Imposition strength levels}
Various different imperatives indicate different strength levels of impositions: you must immediately $X$, 
you must under all circumstances perform $X$, you must $X$, I request you to $X$, you should $X$, 
you ought to $X$,
can you please $X$ now, can you please $X$,
I would appreciate if you $X$, you are advised to $X$.

\subsection{Neutralism, non-obligationism for impositions}
The view on impositions put forward above may be termed ``neutralism with respect to impositions'',
where  neutrality is meant to replace the connotation of unfairness for imposition that all dictionaries 
indicate.

In~\cite{BergstraB2008} a viewpoint towards promises has been worked out that was termed 
non-obligationism. This view implies that a promise need not be characterized by its promissory obligation.
As a stronger view on the independence of the concept of promises from obligations, 
strong non-obligationism was put forward as the viewpoint that the concept of promises may be 
introduced without making any use of the concept of an obligation. Non-obligationism 
being somewhat problematic in certain examples, restricted non-obligationism was put forward as the
viewpoint that for a large class of promises, sufficiently large to be of vital importance for the coordination 
of multi-agent systems, (i) obligations are not needed as a foundational basis, 
(ii) promissory obligations need not characterize the essence of a promise, and (iii) promissory obligations
may in turn be explained in terms of combinations of promises.

Similarly non-obligationism in the case of impositions amounts to the viewpoint that an imposition need not
 be characterized by its impository obligation.

\subsection{Examples of impositions}
Here are some examples of impositions (from $A$ to $B$) that make perfect sense among autonomous agents. It is reasonable that the imposition comes along with some motivation. In many cases an 
imposition can alternatively, though not always more convincingly, be understood as a conditional promise. 
\begin{enumerate}
\item You must pay 50 BTC on Bitcoin account $X$ within one week (that is before date $d_1$), 
otherwise your web sites (with addresses $w_1,w_2$) will suffer
a DDoS attack launched from 10.000 bots for the duration of two weeks, starting at $d_1$.
\item You must pay 10.000 EUR on account $Y$ before date $d_2$ to settle the debt caused by event $E$.
\item You must push the third button from above (as part of a protocol for entering a secured site).
\item You must now take the return money from the cash register outlet.
\item In the coming week you should issue a formal request for reimbursement of your travel costs
of last month (so that the money can be transferred to you in time).
\item You must be careful not driving too fast because police in watching closely 
a few kilometers from here.
\item You must not take the usual way to your work in order to avoid a massive traffic jam.
\item You must be home at 8.00 PM when dinner starts (our guests arrive at 7.30 and they
must leave around 9.30 PM, so please be on time).
\item Please send us your name and the usernames and passwords for your gmail accounts so that 
we can help you to improve the structure of the classification of your email history. (We are well-known
service providers for people having difficulties with dealing with too much email; 
please check our credentials on the following site).
\end{enumerate}

\section{Promises and impositions as instances of Directionals}
We will use the term directional to indicate a directed communication between two agents within a given scope
consisting of the originating agent, perhaps the target agent, and zero or more other agents. Promises and
impositions are classes of directionals.\footnote{%
Searle's directives and commissives will both qualify as subcategories of the directionals.} 
\subsection{Suggestions, proposals, warnings, and predictions}
Besides promises and impositions we will distinguish four more classes of directional utterances.
\begin{description} 
\item[Suggestion:] an option for a course of action or of a state of affairs to be achieved which is 
issued by $A$ to $B$. A suggestion expresses that $A$ has in mind some 
actions or sequence of events, or state of affairs, which $A$ assumes to be possible or reachable for $B$ 
and which $A$ expects $B$ to contemplate as an option. A suggestion of $A$ to $B$ may or may not be
effectuated by $B$.
\item [Warning:] a warning issued by $A$ to $B$ is a suggestion from $A$ to $B$ the 
effectuation of which $A$ considers to be not 
fruitful either form its own perspective or from $B$'s perspective.

\item[Proposal:] a suggestion from $A$ to $B$ the effectuation of which $A$ considers to be 
fruitful either form its own perspective or from $B$'s perspective.
\item[Prediction:] a suggestion (from $A$ to $B$)  that $A$ considers likely to occur, 
irrespective of $B$'s behavior. Predictions encode $A$'s knowledge about the environment and may be used
to transfer that knowledge to $B$.
\end{description}

The idea is that predictions can be used to convey reasoning patterns to other agents. Such reasoning 
patterns can be helpful for agents that must determine expectations generated from promises. Here is an example.

A suggestion from $A$ to $B$ may induce the occurrence of a proposal from $B$ to $A$ which in turn 
gives rise to a promise from $A$ to $B$ that $A$ will cooperate with $B$ when $B$ tries to carry out its proposal.
Subsequently $B$ may promise to $A$ that it will try to carry out the proposal and that it will make use of 
$A$'s last promise.

In Appendix \ref{appendixA} we provide an artificial example involving a large family of promises, proposals, and suggestions.

\subsubsection{Trust maintenance for directionals}
Each directional will have primary side effects on the target agent's state of cognition (mind) and 
secondary side effects on the level of trust that the target agent has in the originating agent. Such side effects are also
plausible for other agents in the scope of the directional. 

Primary side effects are moderated by the accumulated side-effects on trust levels of previous events. The foremost
role of trust is to support target agents (and other agents in scope) of a directional to perform 
expectational reasoning and planning upon taking notice of a directional. Impact on trust levels will usually
take place after assessment has been made of the degree to which the directional has proven helpful or reliable.

\subsection{Voluntary co-op methods, a super-class of Directionals}
Promise, imposition, warning, proposal, suggestion, and prediction, each qualify as methods in the  sense of
object oriented programming, to be 
applied to a target agent by a source agent. As such these are super class (in an object orientation style class hierarchy, though a subset in a set theory style class hierarchy) of the conceivable directionals (which may
also include praise, criticism, and signaling excitement or boredom etc.). 
Because the target agent is always assumed to operate in a 
voluntary fashion upon being influenced by one of these methods, these methods together constitute the category of 
voluntary cooperation oriented agent coordination methods, which we will refer to as voluntary 
co-operational methods, or more briefly as voluntary co-op methods. Below we will often speak of directionals 
instead of voluntary co-op methods.

\subsection{Promise dynamics: central life-cycle and peripheral life-cycle}
Promise dynamics primarily refer to how a single promise moves through its life-cycle. 
More distantly, promises interact in complex ways as agents maintaining various promise bundles
related to different threads of activity may generate new promises, or rather incentives to issue new
promises, as a consequence of reflection upon the existing package of promises each promise
being in its own state of its dedicated life-cycle.

Analyzing the internal activity of agents that triggers their preparation and 
production of new promises is not a 
part of promise theory per se. Rather promise theory provides a language that facilitates system
description while remaining uncommitted to that kind of in depth analysis of individual agent behavior.

The central life-cycle of a promise $p$ indicates that after it has been issued
it persists as a cognition in the minds (memories) of agents involved until on of the following
events occurs:
\begin{enumerate}
\item $p$ is (observably) kept, 
\item $p$ broken, that is demonstrably not going to be kept, 
\item $p$ or withdrawn by the promiser, or 
\item $p$ becomes outdated (faded out).
\end{enumerate}

The peripheral life-cycle of a promise involves modifications of credibility and of trust assigned 
to the promiser, as well as to the promise by agents in scope. The peripheral life-cycle 
also involves acts of determination
of  plausibility (probability, expectation value) of various possible events 
(by agents in scope, and in particular by the promisee) 
given certain trust levels. These plausibilities are the key factor in the reduction of uncertainty
that promising may effectuate.

\subsubsection{A promise life-cycle in more detail}
A promise once issued moves through the stages of a promise life-cycle. Each of the entities of the
entity classes just mentioned moves through a corresponding life-cycle as well.

\begin{enumerate}
\item Promise preparation.
\item Promise issuing and corresponding promise fragmentation and 
distribution.
\item the following steps take place concurrently for all agents in 
scope (each agent taking care of its own
	instance of a promise fragment carrying the local name just mentioned/generated):
	\begin{enumerate} 
	\item promise outcome credibility assessment,
	\item \label{aaa} promiser trust assessment relative to promise outcome,
	\item \label{bbb} promise based expectation generation,
	\item \label{ccc} promise fading update, alternating with promise fulfillment assessment,
	\item repetition of the  steps \ref{aaa}, \ref{bbb}, and \ref{ccc} after each update of 
	promise outcome credibility and promiser trustworthiness, 
	until fading threshold reached or until promise 			
	fulfillment assessment turns positive,
	\item final update of promise credibility and promiser trust,
	\item local (for the agent) promise termination.
	\end{enumerate}
	These steps are carried out by each agent 
	concurrently (that is interleaved for the same agent, concurrently with other agents) 
	with an ongoing reputation production and maintenance (that is exchange and update) process
	performed by each agent (also outside the scope of this particular promise). 
	Reputation updates are
	caused by incoming messages reporting trust modification steps enacted by other agents.
	
	In particular promiser reputation influences trust assessment, 
	which in turn influences promise fulfillment expectation assessment.
\item Global termination of the promise once the last promise fragment (locally) 
carrying its (global but otherwise secret name)  has expired.
\end{enumerate}

Similar life-cycle schemes can be given for other voluntary co-op methods. We we not write out these
matters in detail, with the understanding that these are rather straightforward.

\subsection{Promises and the reduction of uncertainty}
If $A$ promises ($p_1$) $B$ that 
``$A$ is capable of performing an action $c$'', 
that promise may reduce uncertainty for $B$. Indeed upon noticing $p_1$, $B$ knows, 
modulo its trust in $A$ that $c$ is doable.

If $A$ promises  ($p_2$) $B$ with a significant scope $S$ including $B$ that 
``$A$ will not perform action $c$'',
then, if $B$ prefers $c$ to occur it needs to look for other ways, for instance performing $c$ itself.
Clearly $B$'s uncertainty is reduced once more by this second promise.

We assume that some agents in $S$ won't applaud that $c$ takes place, even if 
$B$ is expected to be happy about that event.

Now suppose that $A$ promises  ($p_3$) $B$ that ``$A$ will support $B$ if $B$ performs $c$'' 
with only $B$ in scope of promise $p_3$. 

At once $B$ needs to be very careful. If $B$ fails to notice that the third promise has a very 
small scope, $B$ may judge that an additional incentive (namely the increased support for $A$
after $B$ has performed $c$) has arisen to perform $c$ itself.
If, however, $B$ takes notice of the reduced scope of  $p_3$, $B$ must take the 
possibility into account that $A$ deceives $B$ and will not keep its promise $p_3$ and 
will not show its support after $B$ would perform $c$. 
(Here uncertainty pops up in the form of a potential misunderstanding: 
while $B$ thinks of ``support by $A$'' as being visible to other members of $S$, $A$ may
only think in terms of support shown to $B$ in private.)

At this stage $B$ promises  ($p_4$) $A$ with scope $S$ that ``$B$ will perform $c$ provided that
$A$ promises $B$, now with scope $S$ that it will support $B$ once $B$ has performed $c$''. 
If a counter-promise
from $A$ to $B$ with scope $S$, that  ``it will support $B$ when performing $c$'' is issued by
$A$ then $B$ finds a significant reduction of its uncertainty and may proceed with 
performing $c$ (assuming that support from $A$ will balance opposition from members of $S$). 
Otherwise $B$ has obtained very valuable information:  $A$ may not be trustworthy. 

We find that promises are helpful for reducing uncertainty about 
what can be done, and what will be done and by whom, while at the same time 
the mechanics of promises also creates
new forms of uncertainty, in particular concerning trustworthiness. 
In some cases such forms of uncertainty can in turn be remedied by way of promising.

Certainty as a concept requires much more philosophical analysis that we can provide in this brief paper.
We refer to \cite{Burgess2013} for an account of certainty that is compatible with the aims of this paper.

Besides promises obligations can be a tool as well for the reduction of uncertainty, 
because what is obliged may be likely to happen. Therefore, 
following the line of~\cite{BergstraB2008} we will continue with an analysis of the 
relation between promises and obligations.

\subsubsection{Reduction of uncertainty through other voluntary co-op methods}
An imposition issued by   $A$ on $B$ with scope $S$ (containing $B$) may reduce uncertainty in all 
agents in scope and in particular in $B$ about what $B$ will intend to accomplish. 
It may also reduce uncertainty about which agent will perform a certain task that many agents
expect to lie ahead of at least  some of them. Predictions may reduce uncertainty about an environment.
Suggestions may reduce uncertainty on how to initiate planning, and proposals may reduce uncertainty
about preferences between a variety of suggested options.

A cascade of voluntary co-op methods exchanged between a group of agents may increasingly
reduce uncertainty until each agent feels confident that its plans with be supported by peer agents according
to promises and that occasional impositions will meet an understanding attitude.

A bundle of predictions may set a stage in which a subsequent bundle of suggestions invokes a pluralty
of proposals which in turn are detailed into a network of promises one of which prepare agents for the
exchange of impositions during operation in real time.

\section{Non-obligationism}
In \cite{BergstraB2008} non-obligationism has been proposed as a preferred 
perspective on promises. This means that promises are primarily 
viewed in their capacity as mechanisms for reducing uncertainty and for inter-agent  management
of credibility, trust and expectations.

When promises are used as a method for specification and explanation of artificial agent based
distributed systems obligations need not at all appear, and if only  for that reason a non-obligationist
perspective on promises is profitable because it allows one do do without obligations altogether.
At the same time the management of credibility and trust, as well as the determination of 
quantified expectations for event and states that are sensitive to various promises from an 
existing promise bundle need to be realized by means of sophisticated AI software.

When considering promises as a tool for management science primarily aimed at organizing
distributed human behavior the situation is quite different, on the one hand human agents
seem to have  build in capacities for credibility assessment  and for trust assessment and 
maintenance as well as for the generation of qualified, if not quantified, expectations. 
In addition, however, the existence of obligations, however defined, is a fact of life for human agents. 
Promise theory can contribute to management science by making promises available in a systematic
way based on a non-obligationist interpretation. 
When aiming at a contribution to management science, the interplay between promises and 
obligations requires careful investigation which cannot be simplified by disregarding obligations
entirely.

\subsection{Pseudo-promissory obligations}
Imagine the the following chain of promises:
\begin{enumerate}
\item $B$ offers a service $s$ delivered in units (1 hr sessions at $B$'s office) 
at a price $p$ EUR 
per unit to be paid after successful delivery
of the service. The offer is made as a promise ($r_1$) to a scope including agent $A$.
\item $A$ promises  ($r_2$) $B$ that $A$ is willing to use $2$ units of $B$'s service $s$ and
to compensate $B$ by paying $B$ an amount $2 \cdot p$ EUR within one week after the
final session related to the delivery of this service, 
and upon having received a written (electronic) request for that payment from $B$.
\item $B$ promises  ($r_3$) $A$ to provide two sessions implementing both units of service at
successive times $t$ and $r$.
\item $A$ promises  ($r_4$) to appear at $B$'s office twice at $t$ and $r$ in order to 
consume the successive units of $B$'s service.
\end{enumerate}
Is it the case that any of these promises has engaged $A$ in an obligation to pay $B$? We find that 
there is no such
obligation, instead only actually consuming both units of $s$ engages $A$ in a an obligation to pay.

The obligation for $A$ to pay an amount to $B$ seems to originate from  promise  $r_2$ or 
perhaps from promise $r_4$. It is not directly linked to either of these promises as this obligation
is still somehow conditional.
For that reason it may be called a pseudo-promissory obligation rather than a promissory obligation.
Assuming that it is clear what it means that after having consumed 2 units of $s$ 
at the agreed timeslots $A$ is obliged to pay $B$ and that such an obligation arises in that manner the 
link with promise issuing still is an indirect one only.

This connection between promises and obligations is very common: a conditional promise expresses
that once a condition is satisfied (which requires one or more actions from either parties 
subsequent to the issuing of the promise) that state of affairs creates an obligation. 

In the above example the simplest way for $A$ to understand the obligation at hand is that it 
coincides (consists of) a bundle of promises issued by $B$:
\begin{enumerate}  
\item $B$ promises $A$ that 
after having received the required payment from $A$ (or on behalf of $A$) in due time $B$ 
will not send any further requests for payment connected to that particular episode of 
service delivery from $B$ to $A$, 
\item if no payment is performed by $A$, $B$ will issue another request
adding the cost of so doing plus some amount serving as a penalty, and 
\item
 if, after some fixed period $u$ yet no
payment is made, $B$ will sell (rather than outsource) the cashing of its once more increased 
claim on $A$ at some discount to a third party
(another agent) who will seek to obtain the payments from $A$ on his behalf. 
\end{enumerate}

It seems pointless to ask for a deeper sense of obligation than can be specified by means of
this bundle of promises because concurrently 
$A$ may be complaining about $B$'s poor service and
ask for a promise by $B$ to nullify $A$'s costs or even to 
provide compensation because $A$'s problems
have not been solved but have rather been worsened.

Requests for payment may be understood as impositions, like promises such impositions may be
credible or lacking credibility, stem from an agent that is considered trustworthy to some yet unknown degree,
may be credible and deceptive at the same time and so on. 

\subsubsection{Generation of pseudo-promissory obligations}
In one insists on the production of one or more ``obligations'' as a side-effect of a promise being issued, the
idea of pseudo-promissory obligations is that a promise is supposed to be implicitly extended with one or more
conditional promises in the way exemplified by the case just mentioned.

Thus working with pseudo-promissory obligations involves the application of certain conventions for expanding promises to promise bundles that contain packages of conditional promises representing what is often viewed
as obligations produced by a promise but what is now seen as a special class of obligations that can in fact be 
equated to (or reduced to) bundles of conditional promises.

Looking at the matter form the perspective of obligations rather than from the perspective of promises or impositions 
we are dealing with a special class of obligations which merits some further attention.

\subsection{Conditionally promised promise patterns}
These considerations lead to the following definition of a relevant subclass of obligations: obligations 
the content of which consists of a  pattern of promises that
results as a side effect from issuing a promise. The fact that the pattern arises is likely to be the 
content of previous promises. Such obligations will be called {\em promise pattern based obligations}
or PPB-obligations for short.

As long as one thinks of PPB-obligations a non-obligationist 
understanding of promises provides a consistent viewpoint. Speaking of obligations as shorthands
for underlying promise patterns or bundles may be helpful and efficient. Because meta-promises, that
is promises to (conditionally) issue promises are promises as well, 
promises may create PPB-obligations without contradicting the 
non-obligationist view of promises.

Not all obligations are PPB-obligations, and the interaction between 
promises and non-PPB-obligations provides an area for further research. 
However, restricting attention to PPB-obligations allows for a useful
extension of the non-obligationist view of promises to the practice and 
science of management of human operations. At the basis of the application of promise theory to
management science and practice lies the use of promises that do not create any non-PPB-obligations. 
There seems to be ample room for such applications.

\subsubsection{Irreducible promissory obligations}
Non-PPB-obligations may preferably be called irreducible obligations as the 
reduction of their essence to promises is impossible.
The promise of a witness in court to state the truth and nothing but the truth produces a 
promissory obligation that cannot be reduced to a promise pattern. 
In that sense the obligation is irreducible.

Remarkably this obligation comes about after issuing a promise (or a vow) suggesting that
even in this case somehow promises take priority over obligations.

\subsubsection{Promising without imposing}
Although promises may be understood as directionals that create self-impositions, 
the implicit tenet of promise theory is that promises alone provide a very flexible tool for coordination in a
multi-agent system. Augmenting promises with impositions and other directionals 
is meaningful because it strengthens the expressiveness of the theory
by relieving it from a fundamentalistic focus on promises that seems unnecessary. Moreover, the 
addition of impositions to the picture provides additional clarity about the distance between 
promises and obligations, which can hardly be assessed without first assessing the relation 
between impositions and obligations.

\subsection{Promises and impositions versus decisions}
In~\cite{Bergstra2013a} the approach to decision taking from~\cite{Bergstra2011a} 
(so-called Outcome Oriented Decision Taking, OODT) 
has been contrasted with non-obligationist promissory theory. We recall that in the terminology of OODT 
 a decision is supposed to be taken by a deciding agent and the result of
that action is a decision outcome which specifies what has been decided. In another process a decision outcome
may subsequently be effectuated.\footnote{%
Following~\cite{Goddu2011} decision is subject to a product/process ambiguity, and OODT
incorporates a preference for a process view of decision.} 
In order to have comparable terminology, 
it was suggested in~\cite{Bergstra2013a} that
a promise is issued leading to a promise outcome, the latter being close to what is called a promise body
in non-obligationist promissory theory. 

Now a key difference between a decision outcome and a promise outcome has been identified in ~\cite{Bergstra2013a} as follows: while a promiser is usually expected to be instrumental in putting
a promise outcome into effect (that is keeping the promise), in the case of a decision outcome there is no 
expectation that the decider will be instrumental for putting the decision outcome into effect.

In a similar fashion the difference between deciding and imposing can be understood. For an imposition outcome
to be effectuated it is expected that the impostionee will play an instrumental role, rather than the impositioner. A decision is not targeted to a specific agent. Of course one might contemplate ``decisionary obligations" as being obligations that arise from a decision outcome. Such decisionary obligations are most
plausible viewed as the consequence of preexisting promises about agents being compliant with specific classes of impositions impositions.

To give an example: if the government of $X$ decides to go to war with $Y$ (the decision outcome constituting a declaration intention of war), the effectuation of that decision outcome is based on the military staff
having promised to take into effect at their own responsibility such forms of decision outcomes. 
Once the declaration of war $D_W$ has been produced, the military staff will exchange suggestions, warnings, and proposals, and soon the may issue impositions to their subordinate staff members who in turn will produce
impositions down the hierarchy. The compliance with most of these impositions can be understood in terms of
the impositionee having promised to follow impositions from his/her superiors, assuming a that a 
correct decision taking process lies at the root of such impositions.

\section{Promise features}
Four features of promises were taken into account in our static theory of processes in~\cite{BergstraB2008}.
In this section we will provide a number of additional features for promises and we will propose notations for promises 
allowing to take the additional features into account.
\begin{enumerate}
\item agents, type, and body (taken from~\cite{BergstraB2008}):
	\begin{itemize}
	\item promiser,
	\item promisee,
	\item agents in scope (observing the promise upon being issued),
	\item promise type, promise body,
	\end{itemize}
\item promise issuing coordinates (time, space, phase),
\item promise viewing agent (promise as seen from the perspective of that agent),
\item promise inspection coordinate (time, space, phase coordinates of where is the 
promise looked at by the viewing agent),
\item promise validity interval (to be kept in the interval from time/event/state to time/event/state),
\item promise identification token (an abstract token in pi-calculus style that links 
different agent centered views on the same promise together),
\item promise fading out function (describes the degree of fading out of various 
components of a promise notation).
\end{enumerate}

The features mentioned above are independent of trust based reasoning by agents involved. 
A calculus of trust and credibility needs to be presupposed for promises to be of any use. 

\subsection{Promise statements}
A promise statement is an expression that combines a sample of features instantiated 
for a single promise. 
Promise statements that display information about more features can be developed with ease. Here are some examples.
\begin{description}
\item [Base promise form:] In its simplest form (called ground form), $p [\pi \colon b] q$, 
a promise statement conveys (the name of) a promiser ($p$) a 
promise type $\pi$ a promise body ($b$), and a promisee ($q$). These promise statements,
though with a more figurative notation with an arrow between promiser and promise and promise type and body as a subscript for that arrow, have been introduced  and used extensively in \cite{Burgess2005,Burgess2007,BurgessF2007}. In this notation $p [\pi \colon b] q$ is written as
$p \promise{\pi \colon b} q$.

\item [Scope: ]with $S$ a collection of agents $p [\pi \colon b/S] q$ specifies promise 
$p [\pi \colon b] q$ with all agents in $S \cup \{p\}$ in its scope. 
(Promisee $q$ may or may not be included in $S$.)
\item [Episode:] with $t$ and $s$ instances of time (or other situational 
descriptions from which temporal and or causal ordering  information can be derived) 
$p [(\pi \colon (t,b,s)/S] q$ specifies promise 
$p [\pi \colon b/S] q$ with the additional features that $b$ is 
supposed to be kept after $t$ and before $r$. Thus a promise so specified expires at $r$.
\item [Issuing time:] with $u$ an instance of time: 
$p [u,(\pi \colon (t,b,s)/S] q$ provides the additional information 
that the promise has been issued at $u$.
\item [Observation time:] with $w$ an instance of time (called observation time) 
the promise statement 
$p [w/u,(\pi \colon (t,b,s)/S] q$ provides the additional information that the 
promise statement is considered at time $w$ by an appropriate agent.
\item [Subject fragmentation:] Upon its issuing a promise fragments 
over a community of agents, that is each agent in scope of the promise becomes the 
carrier of a fragment of it. A notation for fragments will include a name $r$ of a
subject (carrying agent) as additional information. Subject $r$ constitutes the 
perspective from which the promise statement is 
considered descriptive of the state of affairs: $p [w,r/u,(\pi \colon (t,b,s)/S] q$ is a promise statement that 
provides the additional information that it is considered, or held, at time $w$ by an 
agent $r$ in $S \cup \{p\}$. 
\item [Subject fragment identification:] For different human agents the common origin of respective 
promise fragments available to them lies in fault prone memories. For artificial agents additional techniques are available, for instance tagging all fragments with a secret key $\alpha$ known to the agents in scope only,
who may use a corresponding public key $\beta$ for calming to an agreement that respective 
fragments have the same promise issuing as an origin. 

By decorating a subject fragment with that key pair an identifiable subject fragment results:
 $p [w,r(\alpha,\beta)/u,(\pi \colon (t,b,s)/S] q$. 

\item [Subject fragment identity binding with alpha-conversion:] In a formal or theoretic account of an agent community making 
use of shared secret keys, instead of a key pair involving a public key, and following $\pi$-calculus 
style process algebra (see \cite{MilnerPW1992}) an alpha-convertible name $x$ 
may be used with in combination with a binder $(\nu x)(...)$. Taking $P$ and $Q$ for names of agents 
and $P[-]$ and $Q[-]$ for contexts formed by these agents in which a promise fragment description 
can be embedded states are denoted by parallel compositions $(P[-]\mid\mid Q[-]\mid\mid... )$. 

Applying the binder and subsequently allowing for alpha-conversion for $x$ one obtains expressions of the following form: \\$(\nu x)(P[p [w,r(x)/u,(\pi \colon (t,b,s)/S] q] \mid\mid Q[p [w,r(x)/u,(\pi \colon (t,b,s)/S] q] \mid\mid ...).$

\item [Fading out:] $p [w,r,F/u,(\pi \colon (t,b,s)/S] q$ adds information 
about the fading out function $F$. This can be explained as follows. 
Upon haven been created when a promise is issued by its promiser, 
the promise statement splits in a distributed collection of subjective promise statements, one for 
each subject in $S \cup \{p\}$. 
Subjective promise statements will fade out and after some time, 
which may extend long after the expiration time of the promise its existence comes to an end. 
A fading out function $F$ can specify at each moment $u$ the degree of fading 
out (being forgotten by the subject) of the components of the promise. 
\item [Fading out for identity carrying subject fragments:]  Subject fragment identification can be 
combined with fading out: $p [w,r(\alpha,\beta)/u, F,(\pi \colon (t,b,s)/S] q$.
\item [Fading out for alpha-converting identity carrying subject fragments:] Fading out can be described in a 
formalized world using alpha-convertible fragment identities:
\\$(\nu x)(P[p [w,r(x),F/u,(\pi \colon (t,b,s)/S] q] \mid\mid Q[p [w,r(x)/u,G,(\pi \colon (t,b,s)/S] q] \mid\mid~..).$

\end{description}

\subsection{Statements for other voluntary co-op methods}
The type $\pi$ need not be a promise type. It can be a type for an imposition or for any
voluntary co-op method, or any directional. A more general type system including types for other 
voluntary co-op methods makes sense and by using typing in that more general way the above description of promise statements can be adapted into a description of imposition statements, proposal statements, warning statements, suggestion statements, and so on.  We will omit the extensive details of this matter.
\subsection{Promise context: concurrent existence of directional fragments}
A promise, and after its split into fragments carried by various agents, a promise fragment
 exists in between of a number (may be zero) of comparable entities. 
The following can be mentioned:
\begin{itemize}
\item other directionals or fragments of directionals, (with the same or with another other issuer),
\item descriptions of fact, descriptions of opinion, descriptions of contracts, 
descriptions of obligations, existing in databases each carried by agents involved,
\item cognitions of fact, cognitions of opinion, cognitions of contract, cognitions of obligation (each supposed to
reside in the minds of various agents involved),
\item wishes, requests, objectives, intentions, plans,
\item states of reasoning arrived at by an agent, that is incomplete sets of conclusions drawn during an 
ongoing inference process.
\end{itemize}

In its full generality the range of possible contexts of promises or other directionals is so 
complex and varied that finding  a general structure theory of such contexts is inconceivable. Clarification of
structure can only be achieved in the presence of simplifying assumptions.
\section{Credibility versus trust for promises}
Once a promiser issues a promise the promise outcome will after some time be assessed by 
the promisee and by other agents in scope according to its credibility. 
That is, given the kind of promiser and the kind of promise outcome, it is assessed by 
agents involved to what extent it is plausible that the promise can be kept. 

If that plausibility is considered very low, fading out of the promise outcome is sped up, 
and the promise may even be terminated without any assessment having been made 
of the promiser's trustworthiness.
In that case it may lead to a negative update of the promiser's trustworthiness even without 
awaiting the time needed to assess whether or not the promise is kept in cases where 
credibility was found sufficient.

\subsection{Credibility assessment mechanism (CRAM)}
An agent noticing a promise being issued will attempt to assess its credibility. 
In doing so the  agent 
applies some kind of credibility calculus to its collection (in memory) of old and new 
promise statements. 
We must assume some mechanism that produces a plausibility that 
the promise can be kept by an agent like the one who issued the promise. 
This mechanism will be referred to as the credibility assessment mechanism (CRAM).

The CRAM takes observations on agent behavior, agent classification, 
and agent performance  as inputs and it produces, 
given a promise statement an expectation of the credibility 
that the promiser will keep the promise taking only into consideration 
the type (class, kind) of the promiser (e.g. a human being will not keep 
the promise to fly like a bird, or to swim across the Atlantic with out 
support, or to walk 100 km without taking food and drinks during the walk. 
For promise lacking credibility the question whether the promiser 
can be trusted is immaterial. However, issuing such a promise 
may decrease the promisers trustworthiness in the eyes of the 
promisee or other agents in scope of the promise. 
The CRAM may make use of a credibility calculus that 
allows to express assessments in rational values between 0 and 1. 

\subsection{Trust assessment mechanism (TRAM)}
Promises that have been assessed as being credible given 
the promiser may still not be kept while other comparable 
promisers may keep similar promises without hesitation. 
Besides credibility trust plays a role. Low trust matters only when high 
credibility has been assessed. Trust depends on the logic
of promiser behavior (if them promisee thinks that keeping a promise
is against promisers self-interest it may lower its trust that the promise 
will be kept). It also depends on past behavior of the particular promiser
and it may depend on the size of the scope, as all agents in scope may
lower their trust of the promiser upon observing that a promise is not kept.
If a promiser values high regard (trust) by the agents in scope that may
constitute an additional incentive to keep the promise.  
The trust assessment mechanism (TRAM) which is operational as a
separate and autonomous functionality for each agent takes observed behavior
as inputs besides a stream of promises. To each promise-promiser pair it can 
assign a degree of trust that the promiser will keep the promise. 

\subsection{Credibility and trust for other directionals}
If someone ($A$) suggests his partner ($B$) to make a financial reservation for the coming 
vacation of an amount of 10.000.000 Euro an immediate clash with credibility may arise.
$B$ may wonder what sort of a vacation might this be about and where is this amount to be found?

Plausibly $B$ may react with a warning to $A$: ``that's nonsense!''. Had $A$ suggested to 
make a reservation of 2500 Euro instead, $B$'s reaction might be to propose $B$
making a reservation of 3000 Euro. If $B$ reacts to this alternative proposal of $A$
with the same warning, that may induce a credibility drop in $A$'s perception of $B$.

Directionals lacking credibility are routed differently from directionals understood by the target
as having adequate credibility. Only in the second case expectational reasoning will take
place and will induce observational activity that may in turn trigger assessments impacting on
trust levels.

\subsection{Promises and self-trust}
Different agents may plan to compute expectations or otherwise quantified plausibilities of future 
events on the basis of promises and trust in the promiser. The very promiser, who may or may not trust itself,
has a special position among these agents. Indeed if $A$ promises ($p$) $B$ with scope $S$
 to perform $c$, a variety of options concerning $A$'s trust in itself can be distinguished:
\begin{enumerate}
\item $A$ may deceive $B$ by having (but not showing) a lack of trust in itself. 
$A$ may even consider itself 
unable to perform $c$ so that self-trust and self-credibility are both very low. This does not imply
that $A$ has a low self-confidence. On the contrary, because $A$ has to deal with adverse 
reaction (degradation of trustworthiness in their eyes) from members of $S$ once $A$ 
breaks its promise, $A$ must be confident that it can deal with that eventual consequence of its promise (and
in particular with the consequences of its expected breaking of the promise).
\item $A$ may have little doubt that it can deliver $c$ and this may be based on information not
available to other agents in $S$ who initially have less trust (than $A$ itself) in 
$A$'s ability to perform $c$ in an adequate manner.
\item $A$ may be overconfident, in which case $A$ is honestly convinced that it will deliver $c$, while
other agents in $S$, who are better informed about the relation between $A$'s capabilities and what is
needed to perform $c$, rightly place less trust in $A$ as a potential actor of $c$.
\item Most agents in $A$ may have high trust in $A$'s capability for performing $c$ but $A$ itself
may not be so sure. $A$ may feel having been put under pressure to issue a promise for doing $c$
that $A$ might have preferred to avoid. 

In this case $A$ may feel deceived by some agents in $S$ who promised to 
make use of a possibly forthcoming  promise by $A$ for performing $c$. 
$A$ may think that these agents should have known that $A$ promising $c$ lacks credibility. 
Perhaps performing $c$ involves certain risks that $A$ prefers to avoid.
\end{enumerate}

The interplay between self-confidence, self-credibility, and self-trust can be very complex. This
complexity is real, however, and promises seem to provide a useful method for dealing with it.

\subsection{Quantification of outstanding promises and impositions}
Each agent may issue promises repeatedly. Two promise issuings convey the same
promise if keeping one of the necessarily implies keeping the other as well. In that
case promise statements are called equivalent. As equivalent promises may be
issued at different moments in time the corresponding promise statements need not
be equal. When describing a scene promise statements should be made so detailed
(high promise statement resolution) that equivalence with other 
promises that occur in the same scene can be reliably judged.

If a promise statement is abstract, that is it contains relatively little information about various
features, equivalence with another promise statement may be hard to assess.

Promise repetition, i.e. the consecutive issuing of different but equivalent promises 
may impact (that is
serve as an input for) both TRAM and CRAM of various agents. Whether repeated
issuing of a promise has positive or negative impact on credibility and trust 
depends on the circumstances.
\subsubsection{Imposition drift}
Once an imposition has been internalized by an agent its life-cycle begins and the imposition, or
rather its residual representation, may transform over time. Agents may change their 
view of what has been imposed upon them, including what they have promised themselves. An agent
may invent new promises that it thinks it has made (while it has not), it may forget impositions
created by itself and by other agents. An agent may have its own strategy for fading out 
impositions and finally forgetting about them. These phenomena are captured under imposition drift.
\subsubsection{Imposition portfolio}
An agent may be supposed to maintain a data base with a portfolio of impositions 
that it has received, including
self-impositions resulting from its own promises. For a human agent this portfolio may range 
from  a very formalized and technically well-supported system to a bundle of more or less 
vague memories. An imposition portfolio may have been modified (perhaps compromised) 
by imposition drift of some of its content.

\section{Promise dynamics: examples involving computer program usage}
\label{PDyn}
Promise dynamics has two main aspects one of which we are now in the position to illustrate by means of examples.
What is shown clearly by the examples below is how trust updating, promising and promise keeping are interrelated.

The effect of a promise being issued lies in the behavior of the promisee being compliant with an expectation 
(in the promisee's perception) that has been generated by the promiser upon issuing the promise. The later effect is
mediated by promisee's trust in the promiser. We will exemplify how that may work and also how different promises may interfere provided both are based upon the same trust variable.

\subsection{Promise assessment and trust level maintenance}
We will consider promise dynamics in a context where some impositions play a roles as well.
The examples below involve a single promise type only: 

$\pi = \pi_\alpha(P,U)$ = ``promises about the adequacy of a product $P$ 
as a tool for performing task $U$''. 

As a typical example of a product we will consider a computer program. With $T_p(q)$ we denote the trust that
$p$ has in $q$. We assume that trust is measured on the five level scale $[-2, -1, 0, 1, 2]$ where 0 is neutral,
-2 expresses strong distrust, - 1 express distrust, 1 expresses trust, and finally 2 expresses strong trust.

We will display several threads of activity involving promises and corresponding trust maintenance. 
\begin{enumerate}
\item The first thread illustrates (i) that observation of failure to comply with a promise leads to decrease of trust
in the relevant promiser, and (ii) that with neutral trust (in the promiser) a promisee ignores a promiser's promise:
\begin{itemize}
\item Initial trust state $T_q(p) = 1$,
\item A promise $m_0$ is issued:\\
$m_0 = p[\pi_\alpha(P,U):\mathit{``P~is~adequate~for~task~U"}/\{p,q,r\}]q$,
\item $q$ installs $P$ and prepares for the use of $P$ for task $U$,
\item $s$ imposes $q$ to perform task $U$,
\item $q$ uses $P$ for task $U$,
\item $q$ observes that $P$ fails for task $U$, and assesses that $m_0$ was not kept,
\item $q$ decreases its trust in $p$: $T_q(p) = 0$,
\item A promise $m_1$ is issued:\\
$m_1 = p[\pi_\alpha(Q,V):\mathit{``Q~is~adequate~for~task~V"}/\{p,q,r\}]q$,
\item $q$ refuses to install $Q$ (and by consequence to prepare it for task $V$),
\item final trust state $T_q(p) = 0$
\end{itemize}

\item On the other hand, observation of a promise having been kept, 
induces increased trust (provided an increase is still possible):
\begin{itemize}
\item Initial trust state $T_q(p) = 1$,
\item A promise $m_2$ is issued:\\
$m_2 = p[\pi_\alpha(R,W):\mathit{``R~is~adequate~for~task~W'"}/\{p,q,r\}]q$,
\item $q$ installs $R$ and prepares for the use of $R$ for task $W$,
\item $s$ imposes $q$ to perform task $W$,
\item $q$ successfully uses $R$ for task $W$, and assesses that $m_2$ was kept,

\item $q$ increases its trust in $p$: $T_q(p) = 2$,
\item final trust state $T_q(p) = 2$.
\end{itemize}

\item Interleaving both threads makes sense and may allow further progress. Different interleaving
strategies (see \cite{BergstraM2007}) lead to different outcomes. In the thread below, which results from interleaving
the first two threads, the program $Q$ is used by $q$ for purpose $V$ instead of the refusal of that use by $q$ caused
by $q$'s neutral trust in $p$ in the first thread.
\begin{itemize}
\item Initial trust state $T_q(p) = 1$,
\item A promise $m_0$ is issued:\\
$m_0 = p[\pi_\alpha(P,U):\mathit{``P~is~adequate~for~task~U"}/\{p,q,r\}]q$,
\item $q$ installs $P$ and prepares for the use of $P$ for task $U$,
\item A promise $m_2$ is issued:\\
$m_2 = p[\pi_\alpha(R,W):\mathit{``R~is~adequate~for~task~W"}/\{p,q,r\}]q$,
\item $s$ imposes $q$ to perform task $W$,
\item $q$ successfully uses $R$ for task $W$, and assesses that $m_2$ was kept,
\item $q$ increases its trust in $p$: $T_q(p) = 2$,
\item $s$ imposes $q$ to perform task $U$,
\item $q$ uses $P$ for task $U$,
\item $q$ observes that $P$ fails for task $U$, and assesses that $m_0$ was not kept,
\item $q$ decreases its trust in $p$: $T_q(p) = 1$,
\item A promise $m_1$ is issued:\\
$m_1 = p[\pi_\alpha(Q,V):\mathit{``Q~is~adequate~for~task~V"}/\{p,q,r\}]q$,
\item $q$ installs $Q$ and prepares for the use of $Q$ for task $V$,
\item $s$ imposes $q$ to perform task $V$,
\item $q$  successfully uses $Q$ for task $V$ and assesses that $m_1$ was kept,
\item $q$ increases its trust in $p$: $T_q(p) = 2$,
\item Final trust state $T_q(p) = 2$.
\end{itemize}
\end{enumerate}
These examples of threads of activity illustrate the interaction between trust updating,
assessment and making use of a program which' adequacy has been promised.

\subsection{Trust level dependent reasoning patterns}
In these examples agent $q$ performs reasoning in order to deal with the implications of its trust in $p$.
That part of $q$'s reasoning proceeds according to a collection of rules. Here are some rules that may be used to
describe control of $q$:
\begin{enumerate}
\item If $p$ promises the adequacy of a program $X$ for task $Y$ to $q$ and $T_q(p) > 0$ then $q$ will  
prepare for the use of $X$ (provided that has not been done already).
\item If $p$ promises the adequacy of a program $X$ for task $Y$ to $q$ and $T_q(p) \leq 0$ then $q$ will not 
prepare for the use of $X$.
\item Given task $Y$, if
\begin{enumerate}
\item $q$ has prepared for the use of a program $X$ the use of which (for some task $Y$) has been 
promised (to $q$) to be adequate 
only by $p$, and 
\item $T_q(p) \geq 0$, 
\end{enumerate}
then $q$ will use $X$ as soon as a request for $Y$ is received by $q$.

\item Assuming that for some task $Y$:
\begin{enumerate}
\item $q$ has prepared for the use of a program $X$ the use of which (for task $Y$) has been 
promised (to $q$) to be adequate only by $p$, and
\item $T_q(p) = -2$
\end{enumerate}
 then $q$ will intercept the preparation and unload that program 
(thus blocking its by $q$ use for whatever task).

\item If for some task $Z$:
\begin{enumerate}
\item $q$ has prepared for the use of a program $X$ the use of which (for task $Z$) has been 
promised (to $q$) to be adequate 
only by $p$, and 
\item  $T_q(p) = -1$, and 
\item  $q$ has made use of $X$ before for task $Z$ (without unloading it in between), and 
\item $q$ is requested to perform $Z$,
\end{enumerate}
 then $q$ will use $X$ to perform $Z$.
 
 In case the first two conditions hold and the fourth conditions holds but the third condition fails, 
 $q$ will not use $X$ to perform $Z$ (and may fail to perform $Z$).

\item If $q$ has prepared for the use of different programs for a task $Y$ and is requested to perform $Y$
it will use that program for which the adequacy has been promised by an agent with highest current trust, if such an agent exists. 

\item If several agents have promised a plurality of programs adequate for task $Y$ 
and different programs have been promised adequate by agents with the 
same maximum trust (trust in them of $q$), then upon a request to perform $Y$,
\begin{enumerate}
\item  per program sums of trust levels from different promising agents are compared 
(relevant only if different agents have made adequacy promises about the same programs) 
and the ``best'' program is chosen, and  if this criterion 
fails to discriminate,
\item  that program is chosen about which the most recent adequacy promise (for task $Y$) 
has been issued by an agent
currently enjoying a maxima trust level (from $q$).
\end{enumerate}
\end{enumerate}

Obviously analyzing the validity, consistency, and completeness of this collection of rules, 
or an appropriate variation of it, 
poses a significant problem in itself.
It is reasonable to assume that $q$ experiences a learning curve through which a combination of such rules stabilizes.
Agent $q$ makes use of appropriate informal logic to organize the application of the various rules that underlie
this part of its reasoning.

\section{Concluding remarks}
We have introduced impositions as a second member besides promises 
of the class of voluntary co-op methods, followed by several other
elements such as suggestions and proposals.Voluntary co-op methods are actions or patterns of activity 
which one agent applies in the direction of another agent in order to bring about, enhance, 
facilitate, or invite, voluntary cooperation. Promises are the key instance of voluntary co-op methods. 
Voluntary co-op methods are collected in a more general class of directionals which may go beyond the coordination of voluntary activity.

We have indicated how, in which cases, and to what extent promissory obligations (and impository obligations) may be 
understood as patterns of obligations which are supposed to be automatically co-generated with promises or impositions.

Then we have extended the notational format for promises know from previous work to include may aspects that
enter the picture when contemplating dynamic aspects. These extensions are generic in the sense that similar 
notations may work for other directionals.

In Appendix A we provide examples of stepwise development of promise bundles and in connection with the coming
about and effectuation of a plan of an agent to buy an item from another agent. The example indicates the relatively
large number of promises, and to a lesser extent impositions, that may occur in the context of a simple plan involving a few 
actions only.
  
In Appendix B we provide an example in human machine interaction where a range of promises, and to a lesser extent
impositions, constitute an essential component of the explanation of system behavior in a context 
with autonomous agents. The example indicates that the language of promises is indispensable for the description
of some human machine interaction scenarios.

In Appendix C we carry on with the example on program usage and the side effects of trust updating. In spite of 
the open ended complexity of the topic, mapping out plausible mechanisms and combinations of mechanisms 
proves to be doable and informative.

\appendix
\section{An artificial example of a promise bundle}
\label{appendixA}
Many examples of promises can be given. In this Section a fixed running example will be used to 
illustrate the relative abundance of promises compared to real actions which the promises may be about. 
The example is artificial in that it has not been derived from a real case.

The example illustrates first of all that promise bundles 
linked to a single and simple activity  can be quite large, 
in addition it becomes obvious that most of the promises must fade out rather quickly
in order to avoid unmanageable agent states.

The example also illustrates the use of some other directionals.

A coherent bundle of promises is involved with a single transfer of an amount $m$ by agent $A$
to agent $B$ in compensation for a service or good S/G that $B$ delivers to $A$. 
\begin{itemize}
\item A proposal $p_1$ by $B$ (with $B$'s management $M_B$ in scope)  
to $A$ to deliver S/G against compensation $m$.
\item A promise $p_2$ issued by $A$ (with $A$'s manager $M_A$ in scope) to $B$ to 
accept S/G, ($p_2$ is a counter promise, also called a promise to use, for $p_1$).
\item A suggestion $p_3$ (issued by $A$) to $B$ ($M_B$ in scope) 
that $A$ is able and willing to pay $B$, as a compensation for S/G,
via a particular informational money, say IM (see~\cite{BergstraL2013a,BergstraL2013b} for informational monies).
\item A proposal $p_4$ by $A$ to $B$ to pay $B$ by way of a specific 
informational money (say IMX).
\item A promise $p_5$  by $B$ to $A$ accept an IMX payment, ($p_5$ a counter promise to the proposal $p_4$).
\item A promise $p_6$ by $B$  ($M_B$ and $M_A$ in scope) to $A$ a to confirm a 
payment of amount $m$ (made by $A$) after it has been received via an IMX channel by $B$.
\end{itemize}
These promises coexist during a single transfer scenario, and in order to understand the role of each promise a
detailed analysis of its dynamics is needed. Each promise evolves though a life-cycle. 
For instance $p_1$ may disappear at once if it is considered not credible to $A$ or to
 $M_B$ that $B$ can deliver S/G. When $p_1$ is considered a credible promise in principle, 
 given what is known about $B$ in most general terms, $A$ will evaluate its trust that $B$ can deliver
 S/G against compensation $m$. The degree of trust in $B$ may be a function of past 
 observed behavior of $B$ by a community of potential clients who maintain reputation based trust calculus about a number of agents including $B$. Updating trust (of $B$) when a 
 promise (issued by $B$)  is kept or not kept requires a sound assessment of promise keeping. 
 That in turn requires that promises need to be wrapped in time intervals and similar constants that
 enable reliable assessment at some moment in time.
 
 \subsection{Promises about motivation, preferences, and activity planning}
 Many more promises may be contained in the promise graph surrounding a single transfer. 
 Here are some promises that may precede $p_{1-7}$. These promises concern 
 $A$'s motivation and preferences. 
 Such matters may be case as promises from $A$ to itself with other agents in scope.
 
 \begin{itemize}
 \item A promise $p^m_1$ issued by $A$ to $A$ with the content that 
 $A$ will be satisfied upon acquiring S/G in exchange of compensation $m$ or below.\footnote{%
 Having issued that promise $A$ can assess its credibility as well  as its trustworthiness. 
 As a part of promise dynamics $A$ may initially deem $p^m_1$ credible but upon further reflection
 $A$ may have limited trust in its truth.}
 \item A promise $p^m_2$ issued by $A$ to $A$ with $B$ in scope with the content that 
 $A$ will be satisfied upon acquiring S/G in exchange of compensation $m$ or below. 
 \item A promise $p^m_3$ issued by $A$ to $A$ with $M_A$ in scope that $A$ currently 
 prefers transferring amounts via informational money IMX  to other means of money transfer. 
 \end{itemize}
 Another collection of promises relate to the way in which $A$ will interact with 
 $B$ when preparing the transfer of S/G.
 \begin{itemize}
 \item A promise $q_1$ issued by $A$ to $B$ that $A$ will visit $B$ at time $t$ and location $l_B$ 
 with the intent to be informed by $B$ about the specifics of S/G,
  \item A promise $q_2$ issued by $B$ to $A$ that $B$ will receive $A$ at time $t$ and location $l$ 
 with the intent to  inform $A$ about the specifics of S/G,
  \item A promise $q_3$ issued by $B$ to $A$ that $B$ will not deliver S/G to any other agent if that
  stands in the way of delivery of S/G to $A$ until 24 hours after $A$ completed its visit to $B$,
  \item A promise $q_4$ issued by $B$ that a car (driven by $A$) can be parked for free
  at $B$'s site 
  when $A$ announces his/her arrival   at the gate of $B$'s premises at or after some time $t-u$ 
  reasonably in advance of $t$ (with $u$ equal to, say, 30 minutes).
  \item A proposal $q_5$ issued by $B$ that a price for S/G will be fixed turing the visit and that the
  offer for S/G against that price will stand for 10 days.
 \end{itemize}
 Some promises connected with the (preparations for) the transfer involve $A$'s partner $P_A$.
 \begin{itemize}
 \item A suggestion $q_6$ issued by $P_A$ to  $A$ that $A$ will use $P_A$'s car 
 provided that it will be returned in time. 
  \item A promise $q_7$ issued by $A$ to  $P_A$ that the car will be 
  returned at time $t+r$ (at location $l_A$) at the latest for subsequent use by $P_A$.
  \item A promise $q_8$ issued by   $P_A$ to $A$ that the car will be ready for 
  use (by $A$) at time $t-s$ (with $t-s< t-u$) at the latest (and at location $l_A$) for 
  subsequent use by $P_A$.
  \item A promise $q_9$ by $A$ to $P_A$ that, after returning from the visit to $B$,
  $A$ will subsequently seek $P_A$'s opinion before promising $B$ to 
  buy/use S/G against compensation $m$.
  \item A promise $q_{10}$ by $P_A$ to $A$ that once having provided a positive  opinion about
  the use/acquisition of S/G, $P_A$ will support $A$ to keep the promise
  to provide compensation $m$ to $B$ upon delivery of S/G.
  \item A promise $q_{11}$ by $A$ to $P_A$ that $A$ will only transfer $m$ to $B$ after
  adequate delivery of S/G.
  \item A proposal $q_{12}$ by $P_A$ to $A$ that $P_A$ will be reachable (for $A$) 
   by phone during    $A$'s visit to $B$.
 \item A promise $q_{13}$ by $A$ to $P_A$ to make use of proposal $q_{12}$.
 \end{itemize}
 Apparently a rather formidable bundle of promises, proposals, and suggestions may constitute the context for a 
single money transfer from $A$ to $B$. The dynamics of each of these promises may impact on the
very occurrence of the deal between $A$ and $B$ and the corresponding transfer.

\subsection{Possible extensions of the promise bundle}
In practical circumstances a bundle of promises connected with a single activity 
can be far more complicated than the example just given. For instance this specific 
example may be extended in various directions:
\begin{enumerate}
\item $A$ may not have a driving license and she may 
wish her daughter $D_A$ to drive the family car. 
This which may be communicated though several promises 
to drive her to $l_b$ and back.
\item $A$ may not be able to comply with a previous promise to $P_A$ about doing some
housekeeping work and an other arrangement may be agreed upon, that agreement being
encoded in an appropriate collection of promises issued in advance of $A$ moving towards $l_B$.
\item $A$ may wish support of $P_A$ in acquiring S/G and may try to arrange that support in 
terms of specific promises issued by $P_A$ that (s)he will help $A$ with the use of S/G if that
might be needed.
\item $A$ may agree with $M_A$ upon a strategy for negotiation with $B$ about different versions
of S/G against different prices and conditions. 
This agreement may again materialize in a collection of appropriate 
promises from $A$ to $M_A$ and conversely.
\end{enumerate} 
So it seems that up to 50 promises may easily 
be involved when planning a single transaction for buying
some good or service.\footnote{%
it seems obvious that most of these promises cannot produce obligations because otherwise 
the complexity of the whole setup explodes. In fact the setup must be somehow robust against 
multiple promises not being kept, and promise redundancy may be vital for plan reliability.} 
The need for an understanding of promise dynamics is obvious from the need to forget about the majority of these once their role has come to an end. Instead of logging all promise descriptions
agents will maintain trust about one-another.

\section{Case study: a recurring  parking exit problem}
\label{appendixB}
The use of promises can be unavoidable in practice. Here is a realistic case study where promises
arise time and again and the main agent has no other option than to deal with a growing 
bundle of related promises.

We imagine a parking lot, say P7 on an industrial site with parking areas numbered P1 to P10.
Agent $A$ works in company $C_A$ and has been issued a particular subscription card with the following 
virtues and features. The parking areas are operated by a company $C_P$ licensed to do so by the
local municipality.

\begin{enumerate}
\item The subscription is called a reduced parking price subscription. It can only be issued by $C_P$
via employers $C$ who provide the subscriptions to there staff members.
\item A staff member of a company $C$ pays an annual fee (say 50 EUR) to $C$ and 
obtains a card in return. The card  is named a ``reduced parking price card (RPPC)''. 
The card provides entry and exit (under certain conditions)  to a subset of the 
parking areas P1,..,P10 which is made known to the employee via an email, shortly before the card is
physically handed over by one of $C$'s support staff members.
\item The standard parking cycle for $A$ works as follows:
\begin{enumerate}
\item $A$ approaches the entry of P7 and hold his RPPC close to a black square on the surface of a 
piece of hardware next to the road and in front of the (entry) barrier.
\item The barrier opens and $A$ drives into P7.
\item $A$ searches (and is guaranteed to find) and empty place and parks.
\item $A$ leaves the car and then walks into his office.
\item When time to leave has come, $A$ returns to the parking are P7, and approaches the 
pay station machine. 
Then $A$ holds the RPPC in front of the machine close to a dedicated area for RPPC's and a price is announced on a little screen.
\item Now $A$ must pay. That needs to be done electronically and there are three options: 
a debit card, a credit card, or a cash card (so-called Chipknip). A selection must be made, and is made
though a simple and well-known interface.
\item $A$ pays and receives an indication that this has succeeded.
\item $A$ returns to the car and drives inside the parking area to the exit stopping in front of the exit barrier.
\item $A$ holds the RPPC in font of a dedicated black area and the exit barrier opens. 
\end{enumerate}

\item Rules of the game. Users of an RPPC are explained the following rules of engagement.
\begin{enumerate}
\item Parked cars must always exit within 48 hours, after that period the issuer has the right to 
withdraw the RPPC.
\item Reduced price is active only between 6.00 in the morning and 23.00 in the night.
\item Cars must be properly parked in demarcated locations (quite hard to see in practice).
\item If there is no free space cars are not admitted until place has again become available, 
that is until other cars have left.
\item Payment of the initial fee provides no guarantee that a free space can be found.
\end{enumerate}

\item Other aspects of the user interface of the parking area equipment.
\begin{itemize}
\item When entering the parking area one always
 has the option to push a button and to receive a paper ticket and to park at 
full cost. This option is open to the public at large.
\item At the entry station and the exit station and at the pay station one finds:
\begin{itemize}

\item A display that allows some 250 characters of text for messages about the state of affairs. (The display
at the pay station is not identical to the one at both other stations).
\item A button that one can push and which is supposed to provide an intercom connection to an
operation room where a staff member working for or on behalf of $C_P$ can answer questions, and may
provide some help in case of complications.
\item A text with a telephone number that may be called in case of problems.
\end{itemize}

\item Several TV cameras provide the control room staff with information about 
what is going on at entrance, exit,
 and pay station.
\end{itemize}
\end{enumerate}
\subsection{Unproblematic complications} 
Here are some minor difficulties that may occur with the use of RPPC 
together with the way in which 
these may be handled.
\begin{enumerate}
\item If one chooses a way of paying which does not work then after some (very clumsy) interaction 
at the pay station one may opt for another mode of payment. 
This is important because each mode may fail for different reasons that may be out of control of a user.
\item If one is refused access with RPPC and is willing to take a paper ticket and pay the full prices, one will
be able to park (provided there is free space). For longer periods that is very expensive, however. 
For short visits it may be a realistic option.
\item If when approaching the parking area the barrier is open and the system is out of action, 
say for maintenance purposes, one may enter and park. 
When exiting under the same conditions one has parked for free and no one complains. 
When exiting some interaction through the intercom system will suffice to convince the 
control room staff that remotely opening the barrier is the most 
reasonable way to proceed and one has parked without paying.
\item If when approaching the parking area the barrier is open, one should try to 
check in with the RPPC but there will be no feedback as to whether that has succeeded.
\end{enumerate}

\subsection{A problematic complication: parking exit problem case I}
\label{pc1}
A problematic complication is a problem for which the user when confronted with
the problem for the first time has no standard way of 
resolving and the impact of which may be hard to assess. 


\begin{itemize}
\item  Either the the pay station display states that it is currently out of order, or 
\item  all modes of payment (open to a user) fail for a variety of reasons 
none of which are in control of the user.
\end{itemize}
At this stage we imagine that $A$ is confronted with the simplest case: 
the pay station display indicates a malfunction. 
It is also imagined that this difficulty arises for the first time (for $A$) so that $A$ 
must now find out how to deal with the matter in an orderly fashion. We imagine that $A$ 
 plans to leave for a lunch at a friday say 12.00  and to return at 14.00. 
When leaving, together with a guest, he finds out that 
the pay station says of itself that it does not work.
Here is a trace of actions of $A$ following that unfortunate event.

\begin{enumerate}
\item $A$ tries to pay and finds that the paying machine does not work, consistent with
its announced self diagnosis. (The machine has promised to be out of order and that 
promise has been kept.)
\item Then $A$ chooses to push the button on the pay station in order to make a call to 
the control room staff. 
(The button represents an implicit promise that such communication can be achieved after its use.)
After some 30 seconds of waiting a staff member, say $Q_{cr}^1$, responds 
(the implicit promise has been kept) and $Q_{cr}^1$ asks for the reason to 
push the button. $A$ explains that the pay station is out of order and that exit with is RPPC 
is impossible for that reason.
\item $Q_{cr}^1$ suggests $A$ to drive the car in front of the exit barrier and to 
call him once more from the intercom next to the exit barrier. (An implicit
promise that the second call will connect to $Q_{cr}^1$ again.) 
$A$ agrees (that is promises to do so) 
and the verbal interaction through the pay station is ended.
\item Once in front of the exit barrier $A$ proceeds with solving the problems as follows:
\begin{enumerate}
\item $A$ pushes the button, waits for some 30 seconds and gets 
connected to the same control room staff member $Q_{cr}^1$.
\item $A$ briefly reexplains the problem.
\item $Q_{cr}^1$ states that he will remotely open the barrier.
\item $A$ proposes $Q_{cr}^1$ to check out his RPPC in ``the system'' 
so that it is known (to the system) that $A$ has exited the parking area (which will allow  a subsequent entry).
\item $Q_{cr}^1$ agrees and the barrier opens, 
$A$ drives forward and is satisfied  that the problem now has been solved.
\end{enumerate}

\item When returning after lunch, arriving with the same guest in the car, 
$A$ approaches the entry barrier and holds the RPPC in front of the dedicated area expecting the 
barrier to open automatically just as it usually does. 
Unfortunately nothing happens and the display indicates that a car is 
already inside  the area (supposedly via the same RPPC)  so that entry is impossible. 

$A$ proceeds as follows:
\begin{enumerate}
\item $A$ pushes the intercom button waits for contact, and is connected with
staff member $Q_{cr}^2$ to whom $A$ explains the problem, now with other cars waiting behind him for entry, and $A$ is told by $Q_{cr}^2$ that he should now take a paper ticked and that when exiting the payment at a reduced price can be made with the RPPC as well after showing the ticket. 

In addition $A$ is told by $Q_{cr}^2$ that he has probably exited the area when the barrier was open forgetting to check out. This hypothetical cause of the problem is denied by $A$ (though to some extent it is true, checkout was not forgotten but rather it was impossible).
\item $A$ takes the paper ticket, the barrier opens and $A$ enters P7 and easily 
parks on a nearly empty P7.
\item At 5.30 PM $A$ proceeds to  leave again (now without guest) 
and $A$ notices that reduced price payment is impossible with the 
combination of the paper ticket and the RPPC. Moreover $A$ finds out from the display that exit is possible when paying full price against the paper ticket (now 18 EUR) or when paying 
73 EUR against the RPPC. 

The state of affairs can be phased convincingly in terms of promises as follows. In response of $A$'s actions the parking system has produced two promises:
\begin{itemize}
\item Upon entering the paper ticket in the pay station and paying 18 EUR the paper ticket will be
returned in a state where it allows exit (when offering the paper ticket at the exit  unit) 
within a reasonable time.
\item Upon showing the RPPC and paying 73 EUR the system ail allow exit (when showing the RPPC
at the exit unit).
\end{itemize}
$A$ chooses not to make use of either promise.
\item Expecting to pay some 3 EUR at most $A$ dislikes both options and $A$ 
pushes the intercom button at the payment station in order to find out how to proceed. 
After talking with staff member $Q_{cr}^3$ $A$ is suggested to drive towards the
exit barrier and to reopen the interaction form there. $A$ promises to do so and after having kept 
that promise, $A$ succeeds to
convince (the same) staff member $Q_{cr}^3$ that the barrier must be opened and that 
he (that is his RPPC) must be checked out.
The barrier opens and $A$ leaves P7.
\end{enumerate}

At this stage $A$ understands that two promises that were issued by control room staff 
have not been kept: 
(i) he has not been checked out
when exiting last time, and (ii) the promise that he could profit from a 
reduced price after taking a paper ticket was unwarranted. $A$ comes to the following conclusions:
\begin{itemize}
\item $A$ becomes aware that he needs a conceptual model of the control room
including a perspective  on the expertise and capabilities of its various staff members.
\item $A$ understands that intercom connection with the control room staff may not suffice to solve these problems. 
\item $A$ concludes that the next time he will phone the indicated phone number immediately 
after entering the parking area and that he will take a paper ticked if that turns out to be needed.
\end{itemize}

\item On the following monday $A$ enters and exits once more after the same kind of discussions 
with control room staff. Now (after having communicated with two more control room staff members
$Q_{cr}^{4,5}$) 
he has found out that:
\begin{itemize}
\item Control room staff cannot consistently answer the question whether or not they can see from their
location that the payment machine has declared itself out of order, 
though they think of themselves that they can see this by means the 
TV camera system while agreeing that resolution is insufficient to read from the display by 
means of the TV image.
\item Control room staff states that they have no expertise about parking cards, that is not a part of their job.
\item Control room staff cannot check in our check out cars 
(wrt. the database of parked cars that underlies RPPC). 
They can only open and close the barrier and they can make 
the machine near the entry produce a paper ticked even if it has not 
been requested by the client in front of the barrier.
\end{itemize}
\item On Tuesday and Wednesday $A$ commutes by means of public transportation.
\item On Thursday, with the third entry after these problems have started $A$ speaks through the intercom 
at the entry barrier with staff member $Q_{cr}^7$ and is explained that $Q_{cr}^7$ 
cannot from there solve these problems, and that 
taking a paper ticket is the only option available at this stage.  
In addition $A$ is told that he must find some higher authority to solve his parking problem with the card.
Thus:
\begin{enumerate}
\item $A$ takes a paper ticket and enters.
\item $A$ phones the number indicated at the entry and gets connected to $Q_{cr}^7$ once more.
$Q_{cr}^7$ complains that it makes no sense to phone him twice for the same issue.
\item $A$ indicates that he could not know that the button and phone number lead 
to the same person and that he saw no other option than to phone the indicated number. 
\item 
\label{pnumber1} $Q_{cr}^7$ reads a $A$ another telephone number (say $N_{bo}^1$)
which will provide access to relevant
back office staff of $C_P$. $A$ promises $Q_{cr}^7$ to contact $C_P$ via  the second phone 
number.
\item Once phoned a person appears who states her name, a piece of information that $A$
forgets. After some explanation $X$ understands the problem, 
and she promises to check out the car (which according to her can be done when the car is
outside P7 as well as when it is inside P7, 
and which will lead to a state from which both entry and exit with RPPC is enabled) 
and states (promises) that from now on things will be back to normal.
\end{enumerate}
\item When leaving that day at 16.00 $A$ finds out that the barrier fails to open 
(the display also shows, and thereby promises on behalf of the parking system, 
that exit can be obtained with the RPPC when paying 125 EUR). He proceeds as follows.
\begin{enumerate}
\item
$A$ drives the car back to a position not standing in the way of other exiting cars and 
once more phones the backoffice number obtained from $Q_{cr}^7$ 
now being connected to another person $Y$ who claims 
not being responsible for P7 and that someone else is in charge, who can be reached under number
$N_{bo}^2$ a piece of information that $A$ forgets. During the same call, however, a
connection to that person, say 
$Z$ is arranged by $Y$ and after having been explained by $A$ the historic account of 
events $Z$ states that he is indeed responsible for P7 and should have been in the loop
in an earlier stage already. In addition he readily admits that he is sometimes is puzzled by the system 
himself  just as well, and that he does not know (though expects) that
once he has checked out the car (a third promise issued but not kept by $C_P$ staff) subsequent
exit will be unproblematic and that $A$ should drive towards the exit and phone him 
once more if it does note work. $A$ promises $Z$ to drive to the exit and to try to 
get out by means of his RPPC.

\item After haven complied with the latter promise, 
$A$ is refused exit and phones the backoffice once more, getting connected with $Y$, 
asks for a connection with $Z$, and $Z$ now promises $A$ that (i) he ($Z$) will 
phone the control room and tell them to open the barrier and that, (ii) subsequent entry and exit
will be normal, and (iii), more generally the problem will have been solved upon exiting P7.
\item $A$ waits 2 minutes, then the barrier opens and $A$ leaves P7. 
$A$ notices that a promise now has been kept kept and $A$ starts trusting $Z$. 
$A$ understands that communication
with $C_P$ staff unavoidably and exclusively 
 leads to promises made by them that may or may not turn out to be kept. Misunderstanding about the
 content of these promises is very likely to occur, while analyzing these in terms of obligations is uninformative.
 \end{enumerate}
 
\item On Friday, one week after the problem appeared, 
$A$ has some doubts about what to do, trust $Z$'s
promise and go by car, or don't trust $Z$'s second promise and take public transportation, 
thus postponing the finalization of the issue to another day. 
That particular Friday is likely to be a stressful day for $A$ in office. From that expectation 
$A$ infers that (the risk of engaging in) extended negotiations with $C_P$ staff must preferably be avoided. 
 
 However, because $A$ knows that entry will be easy
by means of a paper ticket, he opts for the car because its saves a lot of time, accepting the fact that he
may have to pay full price for the paper ticket when exiting if after a stressful day he feels disinclined to negotiate with control room staff from scratch.

\item Indeed (on Friday) entry to P7 is unproblematic with RPPC. Several hours later exit is 
possible against the usual reduced price. $A$ concludes, that $Z$ has kept the second promise
as well, and thereupon $A$ inductively infers that $Z$'s third 
promise has also been kept and that for that reason
in all likelihood the problems have been solved in a satisfactory way, 
so that $A$ can forget all promises that have been issued in connection with parking on P7 
since the first complication arose at the pay station.

\end{enumerate}

\subsection{Trust and credibility}
Credibility plays some role in this example: the statement made by $Q_{cr}^6$ that he 
can see via the TV camera that the display of the pay station indicates that it is out of 
order lacks credibility. But that lack is not clear to $Q_{cr}^6$. 
The statement by $Q_{cr}^1$ that payment at reduced price can be 
performed with RPPC also after entry with a paper ticket lacks credibility 
as the interface of the pay station shows no sign of that option. The promise issued by $X$ that
after resetting (check out, neutralization) of the RPPC, exit is possible and that this step is insensitive to
whether the car is inside or outside P7 lacks some credibility.

Trust plays a role just as well. $A$ notices that once a promise is not kept trust in the promiser
is decreased almost unconsciously and ale markably. Once a promise is issued, 
that seems to create both trust and expectation at the same time. 
Once the promise turns out not to be kept, that is the expectation is proven wrong, trust collapses.

\subsection{Lessons learned for $A$ as a user of RPPC at P7}
Here are some practical lessons that $A$ has acquired from the episode..
\begin{enumerate}
\item Upon entering P7, the simplest understanding of $A$'s action besides physically entering P7
 is this: $A$ promises to make use of an expected forthcoming 
promise to be issued by the parking system, 
to exit at  a reduced price after showing the RPPC and subsequently 
successfully paying the amount due.

However, entering P7 does not engage $A$ in an obligation of any form, 
at least not in an obligations which can be simply and completely stated. Thinking in terms of promises
issued by $A$, by $C_P$ staff members and by the parking system and in terms of implied 
expectations, credibility, and trust, provides a far more flexible and applicable  
model than thinking in terms of obligations.

The problem having been solved coincides with $A$ having the car outside P7 and all 
promises having been discharged, either kept or not kept.
\item Promises issued by $C_P$ staff will necessarily 
play an essential role when solving some complications with the RPPC.
\item When $A$ leaves P7 in an irregular way the probability is high that check out has not 
occurred in a satisfactory manner. That difficulty will not go away and its solution requires contacting $Z$.
That should be done at the earliest convenient opportunity, preferably when the car is outside P7.
\item Control room staff know nothing about the pay station and about the cards and its underlying
information system. But they may not always (or all) be willing to admit that state of affairs. 
They are likely to say whatever ends the discussion without any wish to get it right. 
Control room staff cannot inspect what is on any of the three
displays without the support of the client's visual information gathering on site. 
On the other hand backoffice staff cannot operate the entry and exit 
barriers directly (but they can instruct control room staff to do so).
\item Parking for free seems to be possible for someone who is not afraid of extended and 
repeated intercom discussions, 
and who is not afraid to lie about how he has operated the equipment and about and what is 
shown on various displays.
\item $A$ must distinguish 4 categories of $C_P$ staff:
\begin{itemize}
\item Control room staff, ($Q_{cr}^{1-7}$ in the example),
\item general back office staff, ($X,Y$ in the example),
\item parking area specialized backoffice staff, ($Z$ in the example),
\item on site maintenance staff, (not playing a role in this example, but often active for solving other problems).
\end{itemize}

\item $A$ has no clue as to the scope of various promises. The extent to which discussions with
$C_P$ staff and dimply message histories are logged is unknown to $A$. 
Each category of $C_P$ personnel has their own views, capabilities and competences. These
differences require different styles of interaction from $A$. Very different
levels of theoretical insight in the issues can be noticed. Control room staff seem to assume that
clients like $A$ know in detail what information they can access from their work place. This assumption
is unwarranted (at least for $A$).
\end{enumerate}
Contemplating alternative paths towards the solution of the original problem (pay station out of order) 
several questions remain  for $A$. Answers to such questions matter in view of a potential reoccurrence  of the same problem.
\begin{enumerate}
\item At what times is backoffice staff available? In other words are there times 
of the day when backoffice staff cannot be reached and problems must be solved through
interaction with control room staff only?
\item If the same problem appears once more, what is the most effective solution? Is that
dependent on the time of the day, is it dependent on the time pressure that $A$ is in?
\item Is it possible to take a paper tick at entry (simultaneously with checking in with the RPPC)
so as to have a method available of exiting efficiently (though at higher costs) if the 
same problem arises once more.
\item Is on site maintenance staff able to neutralize an RPPC? Stated differently: is asking for the 
support by on site maintenance staff an alternative for asking for a connection with backoffice staff.
\item Is control room staff able to switch an intercom conversation to backoffice staff so that 
check out can be arranged via the intercom system alone (important when a mobile phone connection is unavailable).
\item Is it advisable for $A$ to find out the answers to the previous 4 questions before the same complication 
arises once more? Or is the probability of these adverse events so low that learning by 
doing suffices in the future as well.
\end{enumerate}

Besides these ``lessons'' there is much room for improvement of the system. Here are some suggestions.
\begin{itemize}
\item Control room staff should be able to read the various displays, and about the barrier status
(probably already visible via TV) and should 
be informed in real time about the pay station status.
\item Once the pay station is out of order RPPC holders must be allowed exit without payment and with
proper check out. (This requires a software modification.)
\item RPPC holders must be able to check out without further payment when the barrier is open. This may require that control room staff visually inspects the situation and validates that a car is at the exit. If the car does not exit and the barrier closes with the car inside P7, control room staff must be able to undo the checkout.
\item If all fails and control room staff must open the barrier while check out of an RPPC holder is in doubt, oral communication of the card number must be possible as a valid form of check out.
\item Control room staff must be instructed on how to communicate validly about RPPCs.
\end{itemize}

\subsection{Aspects of promise dynamics}
This particular case study features an certain mix of promise dynamics. In other examples other 
features may be combined.
\begin{enumerate}
\item
In the parking example quite a number of promises appear, all of which can be forgotten by $A$ (and other agents involved) once the problems have been solved. 
Only a modified trust assignment by $A$ to
various staff member categories results form the episode.
\item 
When a promise is first issued by  $C_P$ staff personnel $A$ assigns a high expectation too its 
being kept and a high trust to the promiser. The very fact that a professional member of the parking 
system staff issues a promise creates a bonus leading to both initial trust and expectation. 

Trust and expectation remain high and unchallenged until either the promise is kept and trust increases or the promise is broken and trust collapses.
\item Reputation based mechanisms play no role in the example. 
Degraded trust of $A$ in parking support staff is turned by $A$ into
a change of the model that $A$ has in mind, thus allowing to deduce that certain promises 
are lacking sufficient credibility to rely on.
Promises lacking credibility are assimilated by $A$ without further degrading their trust in the
issuing promisers because $A$ thinks to understand why (i) these promisers don't know 
how and why not to make such promises, and (ii) that other (credible) promises made by the 
same staff members are likely to be kept, so trust (of $A$) has become agent 
specific and promise dependent.

\item In all circumstances $A$ has been trusted by parking staff to the extent 
needed to resolve acute complications (that is entry and exit). $A$ has no clue as to whether or 
not his handling of the difficulties has modified that trust, and if so for how long and with whom.
\end{enumerate}

\subsection{The parking exit problem and informal logic}
In \cite{BergstraB2008} promise theory has been displayed as a subject in informal logic. This line of
thought merits further contemplation. As it turns out the parking exit problem example provides a 
number of connections to informal logic. 
By first analyzing how informal logic relates to the reasoning that
is applied by various agents without any role of promising it becomes possible to understand some informal logic aspects of promises.

\subsubsection{From induction to deduction}
A classification of reasoning with some support from informal logic (see~\cite{???}) is as follows: deductive reasoning
produces conclusions from assumptions where the conclusions are at least as much 
justified as the assumptions, inductive reasoning produces conclusions from assumptions where the conclusions are plausible (understood in terms of subjective probability) relative to the assumptions, 
and conductive reasoning (or pro and con based reasoning) combines and weights the combined
impact of both supporting and opposing reasons for a single assertion. 
\begin{itemize}
\item
The parking exit problem example provides phases where each of these forms of reasoning
are applied. To begin with the trust that the system will work as intended at any moment of time comes
about from inductive reasoning only. It is certainly impossible for any client to understand al implementation details of the system to such an extent that deductive reasoning provides the certainty that it will operate without flaws.
\item 
Real time reasoning, in particular client based reasoning during (problems with) system use, may
start as inductive reasoning and migrate towards deductive reasoning. The latter takes place once a client starts developing a mental model of the parking system. For instance:
\begin{enumerate}
\item
Consider assumption $R_0$: ``If the pay machine display states `out of order' payments cannot be made''. 

This assumption may first
emerge as a plausible fact in a (learning) phase where a client tries to pay in 
spite of the indication on the display. Then the client may conclude that display status ``out of order'' indicates with high likelihood that
further attempts to use the pay machine are futile until the status has changed. After some time
the client will used deductive reasoning from a mental model comprising rule $R_0$.
\item Assumption $R_1$: ``exit granted by control room staff will not check out the RPPC status''.

Again this fact may be a matter of inductive inference first, only to become an axiom permitting deductive
inference once the client has developed a mental model of the working conditions of control room 
staff (monitoring a plurality of parking areas each dealing with different equipment,
different display systems and error messaging, and with different subscription card policies).

\item Assumption $R_3$: ``when the pay machine is out of order RPC holders need to check out
by way of intercom communication either with $C_P$ control room staff at the exit terminal, or with
backoffice staff via the mobile phone''.

\item Assumption $R_4$: ``control room staff is unaware which problems must be dealt with by
backoffice staff, they only think in terms of sending on site maintenance staff''. $R_4$ is not a consequence of any model of the system, it remains an outcome of inductive reasoning. The
validity of $R_4$ may change in time.
\end{enumerate}
\item Conductive reasoning appears (hypothetically) in several circumstances:
\begin{enumerate}
\item  in the case that $A$ experiences the same
complication once more now after working hours, say at 10.00 PM. 
Now $A$ must determine whether to try to phone backoffice staff first
or to deal with control room staff only and to postpone formal check out to another day.
\item
Conductive reasoning is also called for if $A$ finds the pay station out of order during working hours, 
but at a moment where $A$ must act under severe time constraints.
\item
If during working hours $A$ (for whatever reason known or unknown to $A$) 
 is refused access after showing the RPPC then $A$ needs 
conductive reasoning to determine whether or not entry by means of a paper ticket is to be preferred.
\item 
Conductive reasoning is also called for if $A$ prefers to resolve a refused entry problem by means
of interaction with backoffice staff and $A$ must determine whether or not to allow other persons
to park (by driving away from the entry) before the problem has been resolved.
\end{enumerate}
\end{itemize}

\subsubsection{Induction and conduction on top of deduction}
Once $A$ has developed a model of how $C_P$ operates P7, it becomes possible for $A$ to 
derive the credibility of a variety of promises on the basis of deductive reasoning. 
(E.g. control room staff promising to check out RPPC is not credible). 
Deductive reasoning may govern to a large extent how $A$ will handle a problem.
Then $A$ knows which promises must be viewed in the light of trust management and maintenance. 

Some predictions cannot be made by deductive means and induction remains 
unavoidable in such cases. 
A typical example is that $A$ may assume that control room staff 
cannot connect an intercom exchange to backoffice staff, 
although $A$'s mental model of the system allows for
that option. $A$ has inferred this limitation inductively 
because control room staff does not mention the option. 
But the conclusion might be wrong and might prove wrong when tested explicitly. Another  
example is that $A$ may assume backoffice staff to be unavailable outside normal working hours.
This need not follow in a deductive manner from $A$'s model, but it may follow with some plausibility,
and for that reason it may still be wrong.

Conductive reasoning seems to apply when the same failure is 
experienced in slightly new circumstances. 
It plays a role in plan formation when different priorities or objectives have to be balanced. 
Promise assessment
(in this case) is not a matter of conductive reasoning so it seems.

Credible promises must be considered in the light of promiser trust.
Such promises play a role in inductive reasoning with expectations not only depending on 
promise content but also on the trust $A$ has in the promiser. Reasoning will deliver a quantified plausibility that the promise will be kept. System behavior will  produce assessments as to whether or
not a promise is kept and in either case an update of trust will take place.

Any high expectation that the system can be used via an RPPC as intended (or promised) is 
by necessity the result of inductive inference of some kind. It is unreasonable to expect parking 
client $A$ to apply a deductive reasoning system about this subject which is able to deal with
system failures. Even when dealing with known failures $A$ may need a 
combination of deductive, inductive, and conductive reasoning each 
applied to an approximate model of the parking system and its management practice, and on top
of that $A$ may need both deductive reasoning to assess the credibility of 
promises issued by parking authority staff and inductive reasoning to assess the plausibility that
promises will be kept. The latter form of reasoning taking inputs from a current trust level in
various staff members (or classes) which is updated whenever it comes to light that a promise is kept
or is broken.

\subsection{Parking exit problem case II}
A week later $A$ tried to exit P7 and the pay machine is clearly out of order. $A$ phones the number $N_{bo}^1$
of 
backoffice staff that was communicated before (see \ref{pnumber1} in \ref{pc1} above), and is automatically told 
(promised) that non-one is available
and that after telling name and number the caller will be called back as soon as possible. That happens after
a few minutes, by backoffice staff member, say $Y^{\prime}$, who asks about the problem. Upon 
understanding the cause of the problem and without asking further questions $Y^{\prime}$ connects $A$ back to 
the control room. Now $A$, understands that he will be unable to get anywhere with checking out, and
after some explanation he finds control room staff member (say $Q_{cr}^8$) willing to open the barrier whereupon $A$ exits P7.

Now $A$ still has to check out. Four successive times $A$ calls the backoffice being told by an 
answering machine that his
call will be answered ASAP, which only takes place at the next morning. Now $A$ reexplains the entire chain of events
to some backoffice  staff member, say $Y^{\prime\prime}$, 
who, after asking for the card number neutralizes the card and claims that all will be fine from now.
Subjectively $A$ assesses that three out of the four promises (issued after $A$'s leaving P7)  have not been kept,
but P7 management may claim that (i) by means of a single return call four promises may be kept at the same 
time, and (ii) returning the call the next morning has been the best they could do. By now $A$ regrets not having
written down $N_{bo}^2$ as it constitutes a connection that might yet give access to a living entity.

\section{Promise dynamics continued}
In this Section we will continue with the extensive example from Section~\ref{PDyn} above. 
We will first expand the notation for trust with aggregates that are specific for either programs or application areas.
Then we will survey a plurality of attributes both for programs and for tasks that may need consideration in a more
comprehensive setting. Finally we consider reputation based mechanisms for trust modification.
\subsection{Extending the  trust scale and mechanism}
Of course the trust of $q$ in $p$, denoted by $T_q(p)$ can be measured in a linear ordinal 
scale with higher resolution than five levels. Doing
so without clear examples of its use is less convincing, however. 

Trust maintenance can be described by making of a family of dedicated trust levels rather than a single one.
Here are some examples, still in the context of programs $P,Q, R,...$ supposedly usable for tasks $U,V,W,..$..
We notice that, in the context of the previous examples, $T_q(p)$ may be understood as $q$'s trust in $p$ in its 
capacity of being  a supplier of programs, or in its capacity of being a consultant about programs that have 
been supplied by third parties.

\subsubsection{Aggregates for specific accumulation of appreciation}
Viewing trust and credibility as forms of appreciation, confidence and dependability can 
be categorized as such as well. 
By focusing appreciation concerning specific themes dedicated aggregates of appreciation
can be introduced for the accumulation of findings, judgements, and sentiments. Here are some examples:
\begin{description}
\item[task oriented credibility:] Given a task $U$, $C\!R_q(p,[~],U)$ represents $q$'s view on the credibility of
 $p$ as an agent authoritative on the suitability of a range of programs for task $U$.
\item[task oriented trust:] Given a task $U$, $T_q(p,[~],U)$ represents $q$'s trust in $p$ as an agent authoritative on
the suitability of a range of programs for task $U$.
\item[program oriented credibility:] Given a program $P$, $C\!R_q(p,P,[~])$ represents $q$'s trust in $p$ as an agent authoritative on
the suitability of program $P$ for a range of tasks.
\item[program oriented trust:] Given a program $P$, $T_q(p,P,[~])$ represents $q$'s trust  in $p$ 
as an agent authoritative on the suitability of program $P$ for a range of tasks.

\item[program/task oriented confidence:] Given a program $P$, and a task $U$, $C_q(P,U)$ represents $q$'s confidence
that program $P$ is useful for task $U$. 

Confidence is an abstraction (from promiser's identities) that $q$ manufactures after having been issued promises by one or more promisers about the quality of $P$ in relation to $U$.
\item[program/task matching credibility:] Given a program $P$, and a task $U$, $C\!R(P,U)$ represents a general
level of credibility that program $P$ is useful for task $U$. 

An appropriate level of matching credibility may be found from the documentation of the program.

\item[program/task subjective dependability:] By abstracting from (averaging out over a variety of) 
judgements of individual agents, a subjective
(or rather intersubjective) trust level can be introduced for dependability.

 $D(P,U)$ represents an agent 
community's evidence based perception of the degree to which $P$ is suitable for $U$.
\item[program/task objective dependability:] Rather than taking subjective, though often assessment based, appreciations
as a basis for dependability, objective criteria (testing, verification, validation, software process certification etc.) 
may be taken
as a basis for the development of an attribution of dependability to a program/task pair. 

Such information may be brought into circulation in a reputation flow based trust management network by an agent
with a high status on software quality management and assessment.

\end{description}

\subsubsection{The meaning of aggregate levels, an outline}
For each of these dimensions of appreciation: credibility, trust, confidence, and dependability, 
the same three key questions can be posed regarding dynamics and impact. We will
provide some  provisional answers to these questions:
\begin{itemize}
\item What is a plausible explanation (informal meaning) of a level on the five point scale?
We will only consider the program or task specific credibility and trust. 
	\begin{itemize}
	\item Task $U$ oriented credibility for an agent $p$ (in the eyes of $q$) is plausibly as follows:
	\begin{enumerate} 
	\setcounter{enumi}{-3}
	\item  (low)  if (i) $p$ has 
	no experience with  of advise about the usability of programs for tasks related to $U$, and (ii) $p$ is professionally 
	connected with a program producer offering programs supposedly capable of supporting with task $U$,
	\item (moderately negative) if either $p$ is not independent or $p$ is lacking experience,
	\item (neutral) if $p$ is independent and if in addition $p$ has relevant experience. 
	\item (moderately positive) if $p$ has been consulting on a range of functionalities 
	comparable to but different from $U$, and,
	\item  (positive) if in addition to the virtues creating a moderately positive judgement $p$ has a 
	recognized reputation for 	the task at hand.
	\end{enumerate}
	
	\item Program $P$ oriented credibility for an agent $p$ (in the eyes of $q$) is plausibly as follows:
	\begin{enumerate} 
	\setcounter{enumi}{-3}
	\item  (low)  if (i) $p$ has no experience with  of advise about the usability of the program 
	$P$ or close relatives of it, and (ii) $p$ is professionally 	connected with the producer of program $P$,
	\item (moderately negative) if either $p$ is not independent (from the producer of $P$) or if $p$ is lacking 
	experience with consulting about the capabilities of $P$,
	\item (neutral) if $p$ is independent (in the relevant way) and if in addition $p$ has relevant experience. 
	\item (moderately positive) if $p$ has been consulting on a range of applications of $P$,
	\item  (positive) if in addition to the virtues creating a moderately positive judgement $p$ has a 
	recognized reputation for consulting on applications of $P$.
	\end{enumerate}
	
	\item Task $U$ oriented trust for an agent $p$ (in the eyes of $q$) is plausibly as follows:
	\begin{enumerate} 
	\setcounter{enumi}{-3}
	\item  (low)  if $q$ has been wrongly advised before at least twice by $q$ about 
	programs for task $U$ and the two most recent
	experiences of $p$ with $q$'s advice on this matter were negative,
	\item (moderately negative) if not low and if the most recent experience of $p$ with $q$'s advice was negative,
	\item (neutral) if $p$ has no relevant experience with $q$'s advice on the matter,
	\item (moderately positive) if $p$ 's most recent experience with $q$'s advice was positive,
	\item  (positive) if $p$ had two consecutive positive experiences with the advice of $q$, and these were 
	$p$'s most recent experiences with $q$.
	\end{enumerate}
	This listing is incomplete because a reputation mechanism may overrule $p$'s reliance on its own 
	experience with $p$. When $q$ is told about two other agents having very recent positive experiences 
	with $q$'s advice on the capability of programs to provide support for task $U$ the trust level may be raised
	level may be raised, 	and conversely.
	
	\item Program $P$ oriented trust for an agent $p$ (in the eyes of $q$) is plausibly as follows:
	\begin{enumerate} 
	\setcounter{enumi}{-3}
	\item  (low)  if $q$ has been wrongly advised before at least twice by $q$ about 
	applications of $P$, and the two most recent
	experiences of $p$ with $q$'s advice on this matter were negative,
	\item (moderately negative) if not low and if the most recent experience of $p$ with $q$'s advice was negative,
	\item (neutral) if $p$ has no relevant experience with $q$'s advice on the matter,
	\item (moderately positive) if $p$ 's most recent experience with $q$'s advice was positive,
	\item  (positive) if $p$ had two consecutive positive experiences with the advice of $q$, and these were 
	$p$'s most recent experiences with $q$.
	\end{enumerate}
	This listing is incomplete because a reputation mechanism may overrule $p$'s reliance on its own 
	experience with $p$. When $q$ is told about two other agents having very recent positive experiences 
	with $q$'s advice on applications of $P$ the trust level may be raised
	level may be raised, 	and conversely.
	
	\end{itemize}
\item Which events produce updates of the levels? 

It is implicit in the above descriptions how credibility levels and 
 trust levels may change in the course of the interaction with agent $q$.
\item What effects on the handling of directionals can be expected from the different levels?

Low or moderately low credibility of $p$  in the eyes of $q$ (for consulting service $s$) 
plausibly has the effect that $q$ will not ask $p$ to provide $s$. In the presence of positive credibility
$q$ may have a preference for an equally credible consultant who is most trusted.
\end{itemize}

\noindent An example of the working this machinery may read as follows:\\

In the style of the examples in Section \ref{PDyn} one may imagine an agent $p$ promising to $q$ that 
usability of program $P$ for task $U$, and agent $p^{\prime}$ promising to agent $q^{\prime}$
that program $P^{\prime}$ is suitable for the same task. 

After agent $r$ has issued the imposition on $q$ to 
perform $U$, $q$ looks for an agent $c$ such that  $c$'s task oriented credibility $C\!R_q(c,[~],U)$ 
is sufficient (level 1 or level 2) and such that among its peers $c$'s task oriented trust ($T_q(c,[~],U)$) is
maximal. 

Having found $c$, $q$ proposes that $c$ will consult $Q$ about which of the two programs is best suited
for performing task $U$. Then $c$ may accept the job of consulting $q$ by promising $q$ that it will do so,
by way of issuing a proposal.
Subsequently $q$ promises to make use of $c$'s advice and after having noticed the proposal by $c$ for a choice
between both programs $q$ chooses the program to be applied accordingly.

\subsubsection{Product, task, user, and provider attributes}
In order to extend the examples of the use of refinements of the trust scale in connection with
the issuing of and reaction to directionals by a plurality of agents we extend the setting of programs
and tasks with additional attributes of both. We will assume that each attribute is measured in an ordinal 
five point scale with -2 representing very low and 2 representing very high. 

Below we made an attempt to list a fairly
comprehensive collection of attributes that may arise in the context of our running example. Although 
only a few of these attributes will play a role in subsequent examples, for the remaining attributes
examples of their use in the context of the exchange of promises and impositions can be easily imagined.

\begin{enumerate}
\item program development time,
\item program development cost,
\item program quality (speed, precision, flexibility),
\item program manufacturing process documentation availability,
\item program system/installation/hardware specificity,
\item program testability,
\item program maintainability,
\item program user base size,
\item program size (e.g. measured in LOC),
\item program dependability (1- probability of occurrence of failure during 1 year of normal use),
\item task description availability,
\item task ubiquity (many agents in need of the functionality),
\item task complexity,
\item task safety criticality,
\item task evolution speed,
\item user awareness of required task functionality,
\item user dependency on task,
\item user access to alternative program providers for the given task,
\item user competence for program/task failure detection and diagnosis,
\item provider track record for producing programs for given task,
\item provider size,
\item provider profitability and stability,
\item provider reputation,
\item provider certification,
\item provider software process documentation available,
\item provider software process maturity level,
\item provider software process involves formal specification and verification,
\item provider dependence from the market of given task oriented programs.
\end{enumerate}

\subsubsection{Expanding the trust network and mechanisms}
We will now expand the setting of the example with the assumption that agent (program constructor) 
$C_P$ is the provider of program $P$, that $C_R$ has constructed $R$, and so on, and that this and 
much more information provides a background for all promises and other directionals about $P$.

In the presence of  information regarding these attributes several additional rules of behavior can be contemplated.
In practice a vast and hardly systematically charted collection of such rules may underly the control logic
of agent $q$'s assessment and update of $p$'s credibility, as well as $q$'s manner of making use of 
resulting trust levels.

On the background trust maintenance concerning program providers is needed, 
and its relation with consultants (such as $p$)
must be captured in a suitable logic.

\begin{enumerate}
 \item If $p$ promises $q$ that $P$ is adequate for task $U$, then program oriented credibility of $p$ is low
 and for that reason $q$ may not install $P$ for the use for task $U$,
 provided one of the following (combinations of) conditions is satisfied. 
	\begin{itemize}
	 \item (i) provider size is very low  and (ii) program size is very high, or
	 \item (i) program cost are very low,  (ii) program user base is small, (iii) user dependency on task is high,
	 (iv) task ubiquity is low, and (v) program development time is high, or
  
 	\item (i) program dependability is low, (ii) task safety criticality is high, (iii) 
	user dependency on task is high, or
	\item (i) program maintainability is low, (ii) task evolution speed is high, and (iii) user awareness of required
	task functionality is low, and (iv) user access to alternative program providers for the given task is high.
	
	\end{itemize}
	We notice that many more such combinations of conditions can be found.
	 \item If $p$ promises $q$ that $P$ is adequate for task $U$, then program oriented credibility of $p$ is high
 and for that reason $q$ will install $P$ and prepare it for the use for task $U$,
 provided one of the following (combinations of) conditions is satisfied. 
	\begin{itemize}
	\item (i) program quality is high, (ii) program dependability is high,  (iii) program user base is large, 
	(iii) user dependency on task is moderate, or
	\item (i) program dependability is high, (ii) task safety criticality is high, (iii) 
	user dependency on task is high, and (iv) user access to alternative program providers for the given task is low, or
	\item (i) program maintainability is high, (ii) task evolution speed is low, (iii) user awareness of required
	task functionality is high, (iv) user access to alternative program providers for the given task is low.
	(v) provider reputation is moderate, (vi) program cost are moderate, (vii) program development time is moderate,
	and (viii) task description availability is high.
	\end{itemize}
	Again we notice that many more such combinations of conditions can be found.
\item If (i) $p$ promises $q$ that $P$ is adequate for task $U$, then the task oriented credibility of $p$ is high
and for that reason $q$ will install $P$ and prepare it for the use for task $U$,
provided one of the following (combinations of) conditions is satisfied.
	\begin{itemize} 
	\item (i) provider track record for producing programs for task $U$ is high, 
	(ii) task is of low safety criticality, (iii) user competence for
	program/task failure detection and diagnosis is high, and (iv) task is highly user specific, or
 	\item (i) provider track record for producing programs for task $U$ is high, 
	(ii) task is highly safety critical, (iii) user awareness of required task functionality is high, (iv) user competence for
	program/task failure detection and diagnosis is high, (v) user dependency on task is high, 
	and (vi) task is highly user specific, 
  	\end{itemize}
\end{enumerate}
The supply of such rules seems endless, though in practice a learning system might develop such rules 
(semi-)automatically and add the rules by need to its rule base.
\subsection{Balancing imposition strength and promiser trust levels}
In the examples above we have assumed that a request on $q$ to perform a task $U$ 
takes the form of a corresponding  imposition on $q$ issued by some other agent $s$. Now impositions are 
request for voluntary cooperation, and for that reason $q$'s trust in $s$ enters the picture. Assuming again a five 
level scale for that from of trust one may wonder how it might interact with the scenario's as outlined above. Here
are two rules for the interplay between impostioner trust and product supplier/consultant trust.

\begin{enumerate}
\item Suppose that  $s$ imposes $U$ on $q$, then if
	\begin{enumerate}
	\item $T_q(s) = 2$, and
	\item $q$ has installed and prepared (only) program $P$ for task $U$, and
	\item $q$ has been promised that $P$ is adequate for $U$ only by $p$,
	\end{enumerate}
	then:
	\begin{enumerate}
	\item if $T_q(p) = 2$ then $q$ will use $P$ for task $U$, and
	\item if $T_q(p) = 1$ then $q$ will issue a warning to $s$ that its relevant trust  level is positive but not optimal,
	and wait for a reply by $s$ (either the proposal not to be bothered and to carry on with using $P$ for task $U$, or
	the proposal to quit complying with its previous imposition altogether), and
	\item if $T_q(s) \leq 0$ then $q$ will propose $s$ to withdraw its imposition (for $q$ to perform $U$).
	\end{enumerate}
This rule embodies the idea that $q$'s high trust in $s$ is reflected by $q$ applying maximal scrutiny to avoid $s$
being confronted with a failure when $U$ is performed by $q$. Here $q$ prefers not delivering service $U$ to risking
the delivery of a faulty service.

\item Suppose that  $s$ imposes to perform $U$ on $q$, then if
	\begin{enumerate}
	\item $T_q(s) = 1$, and
	\item $q$ has installed and prepared (only) program $P$ for task $U$, and
	\item $q$ has been promised that $P$ is adequate for $U$ only by $p$,
	\end{enumerate}
	then:
	\begin{enumerate}
	\item if $T_q(p) \geq 1$ then $q$ will use $P$ for task $U$, and
	\item if $T_q(s) \leq 0$ then $q$ will propose $s$ to withdraw its imposition (for $q$ to perform $U$).
	\end{enumerate}

Having less trust in $s$ than in the case of the first rule, $q$ takes a higher risk of failure when performing $U$
with the help of $P$ upon the request by $s$.
\end{enumerate}

\subsection{Reputation infection}
The existence of a community $C_{p}$ of agents, each independently maintaining trust in $p$, suggests
consideration of mechanisms for allowing $q$'s trust in $p$ to be positively affected by the presence of
high trust in $p$ for a significant number of other members of $C_{p}$.

If we define $p$'s reputation within  $C_{p}$ as the distribution of trust in $p$ over members of  
$C_{p}$, then  reputation infection  takes place if reputation evolves to modified reputation by means
of a mechanism which involves comparison and communication of trust levels between different
members of  $C_{p}$ only.

\subsubsection{Letter of recommendation (LOR) based reputation flow}
In this paragraph we will outline how the spreading out of trust and the conversion of trust into 
reputation might work in the context of our running example.

Suppose that $q$ entertains $T_q(p) =0$ about $p$ and a promise 
$m_0 = p[\pi_\alpha(P,U):\mathit{``P~is~adequate~for~task~U''}/\{p,q,r\}]q$ is issued by $p$. Rather than
refusing to install $P$, $q$ may first propose to $q^{\prime}$ a peer of $q$ that $q^{\prime}$ tells $q$ about its
trust in $p$. The reaction of $q^{\prime}$ to this proposal determines how $q$ will deal with $p$'s promise $m_0$.
\begin{enumerate}
\item If $T_{q^{\prime}}(p) = -2$ then $q^{\prime}$ communicates that fact to $q$, and $q$ adapts $T_q(p) $ to 
$\min (-1,T_q(p))$.
\item If $T_{q^{\prime}}(p) = -1$ or $T_{q^{\prime}}(p) = 0$ it is plausible that $q^{\prime}$ refuses 
(that is promises not to) send this information to $q$. 

\item If $T_{q^{\prime}}(p) =1$ then $q^{\prime}$ will communicate that fact to $q$ upon which
$q$ sets $T_q(p) $ to $0$. 

\item If $T_{q^{\prime}}(p) =2$ then that is communicated by $q^{\prime}$  to $q$ 
upon which $q$ sets $T_q(p) $ to $1$.
 In the latter case it is plausible that $q$ reconsiders promise $m_0$ just issued by $p$.
\end{enumerate}

This mechanism involves the request (cast as a proposal) for a letter of recommendation 
(LOR, issued by $q$ to $q^{\prime}$ about $p$). In case $T_{q^{\prime}}(p) \geq 1$ that LOR is produced by
$q^{\prime}$ in the form of an imposition (by $q^{\prime}$ on $q$) to take notice of that state of affairs.

This very simple mechanism of reputation based trust generation can easily be included in the above examples.

\subsubsection{Third party survey based reputation infection}
A third party $u$ may regularly perform polling of a subgroup $C_p^u$ of  $C_{p}$ and compute an average
trust level $T_{C_p^u}(p)$ over $C_p^u$. Agents in $C_p$ who have no recent observations impacting their trust levels may
prefer to reinitialize their trust in $p$ by approximating  $T_{C_p^u}(p)$ from below.

\subsection{Informal logic}
We have not made an attempt to display the simultaneous operation of the various mechanisms. 
Such matters can be easily imagined, however. What will emerge is a large collection of rules comparable to the
rules that we have given above. Complying with a large set of such potentially incompatible rules without systematic
analysis and preparation is difficult. 

A human agent based trust management system operates without systematic description and analysis. 
When, however, promises, impositions, and other directionals are used in the design and operation 
of artificial agents, artificial trust management cannot be avoided. Designing artificial trust management may
involve the observation of comparable human trust management and rule extraction from such observations.
Rule extraction requires inductive logic. Once under way human agents will be needed to complement the 
automated functionality of an agent community in order to improve and extend the rule set. 
Informal logic (see~\cite{Groarke2013,Johnson2006,Pinto2009,Walton2012}) may then be needed to 
allow an agent to develop new rule or to modify rules. These matters ask for a combination of machine learning, 
informal logic, and formal deduction. Investigating such combinations is left for future work.

\end{document}